\documentclass[11pt]{article}
\usepackage[utf8]{inputenc}
\makeatletter
\def\set@curr@file#1{%
    \begingroup
    \escapechar\m@ne
    \xdef\@curr@file{\expandafter\string\csname #1\endcsname}%
    \endgroup
}
\def\quote@name#1{"\quote@@name#1\@gobble""}
\def\quote@@name#1"{#1\quote@@name}
\def\unquote@name#1{\quote@@name#1\@gobble"}
\makeatother
\usepackage{pdflscape}
\usepackage{multirow}
\usepackage{amsmath}
\usepackage{amsthm}
\usepackage[dvipsnames,table,xcdraw]{xcolor}
\usepackage[english]{babel} 
\usepackage[toc,page]{appendix}
\usepackage{amssymb}
\usepackage[centercolon]{mathtools}
\usepackage[textsize=footnotesize]{todonotes}
\DeclareUnicodeCharacter{2009}{\,} 

\usepackage[utf8]{inputenc}
\usepackage[T1]{fontenc}
\usepackage{mathptmx} 
\usepackage[scaled=0.82]{beramono}
\usepackage{tabularx}
\usepackage{authblk}
\usepackage{rotating}

\usepackage{float}
\usepackage[algo2e]{algorithm2e}
\usepackage[]{algorithm}
\usepackage{graphicx}
\usepackage{here}
\usepackage{booktabs}
\usepackage{pdfpages}
\usepackage[onehalfspacing]{setspace}
\usepackage{cleveref}
\usepackage{cases}
\usepackage{lmodern}
\usepackage{multirow}
\usepackage{bm}
\usepackage{url}

\usepackage{csquotes}
\usepackage{ulem}
\newcommand{\argmin}{\mathop{\mathrm{argmin}}}

\usepackage{cite}

\usepackage{color}
\usepackage{ifdraft}
\renewcommand{\r}{\textcolor{red}}
\ifoptionfinal{\newcommand{\delete}[1]{}}
{
	\newcommand{\stkout}[1]{\ifmmode\text{\sout{\ensuremath{#1}}}\else\sout{#1}\fi}
	\newcommand{\delete}[1]{\r{\stkout{#1}}}
}

\makeatletter
\def\tagform@#1{\maketag@@@{[\ignorespaces#1\unskip\@@italiccorr]}}
\makeatother
\usepackage{subcaption}
\usepackage[paperwidth=8.5in, paperheight=11in, margin=1in]{geometry} 

\begin{document}

\title{Dynamic transitions for fast joint acquisition and reconstruction of CEST-$R_{ex}$ and $T_1$}

\author{Markus Huemer$^1$, Clemens Stilianu$^1$, Nick Scholand$^1$, Daniel Mackner$^1$, Martin Uecker$^{1,3}$, Moritz Zaiss$^{2}$, and  Rudolf Stollberger$^{1,3}$\\
\small{$^1$Institute of Biomedical Imaging, Graz University of Technology, Graz, Austria\\$^2$Institute of Neuroradiology, Friedrich-Alexander-Universität Erlangen-Nürnberg (FAU), University Hospital Erlangen, Erlangen, Germany \\ $^3$BioTechMed Graz, Graz, Austria}}

\maketitle

\textbf{Running head:} Dynamic transitions for fast joint estimation and reconstruction of CEST-$R_{ex}$ and $T_1$

\textbf{Correspondence to:} Rudolf Stollberger - rudolf.stollberger@tugraz.at\\

\textbf{Funding:} This research was funded in whole, or in part, by the Austrian Science Fund (FWF) I4870 and F100800. For the purpose of open access, the author has applied a CC BY public copyright license to any Author Accepted Manuscript version arising from this submission.\\

\textbf{Word Count} (approximately) 195 (Abstract) $\sim$5000 (Body)\\

To be submitted to \textit{Magnetic Resonance in Medicine} as a Research Article.\\
Part of this work has been presented at the ISMRM Annual Conference in 2024 and 2025.

\clearpage
\section*{Abstract}
\textbf{Purpose:}\\
This work proposes a method for the simultaneous estimation of the exchange-dependent relaxation rate 
$R_{ex}$ and the longitudinal relaxation time $T_1$ from a single acquisition.

\textbf{Methods:}\\
A novel acquisition scheme was developed that combines CEST saturation with an inversion pulse and a Look-Locker readout to capture the magnetization evolution starting from the inverse transient Z-spectrum.
The corresponding signal model, derived from the Bloch-McConnell equations, describes both the transient Z-spectrum and the Look-Locker dynamics.
A model-based reconstruction approach is employed to jointly estimate $R_{ex}$ and $T_1$. 
The proposed method was validated using a numerical phantom and benchmarked against conventional CEST and Look-Locker $T_1$ mapping in phantom and in vivo on a clinical 3T scanner.

\textbf{Results:}\\
The joint estimation approach demonstrated strong agreement with ground truth and conventional methods across a wide range of $T_1$  and CEST parameters. The acquisition time was reduced by 20-30\% compared to standard CEST protocols, while providing higher signal-to-noise ratio (SNR) in parameter maps.

\textbf{Conclusion:}\\
The proposed technique enables robust and efficient simultaneous quantification of CEST $R_{ex}$ and 
$T_1$ in a single acquisition. It improves parameter map quality and reduces scan time, making it suitable for both phantom and in vivo imaging across a wide range of physiological conditions.\\
\subsection*{Keywords} CEST, T1, AREX, REX, model-based reconstruction, TGV, Look-Locker, QUASS, APTw, ssMT, rNOE
\clearpage
\section{Introduction}
In Chemical Exchange Saturation Transfer (CEST) imaging the contrast depends indirectly on the chemical exchange between the bulk water and protons in solute molecules.
The technique saturates protons using off-resonant radiofrequency (RF) pulses that significantly increases
the sensitivity of the image contrast towards various molecules
compared to direct saturation \cite{vanZijl2018}. 
The spectrum of indirectly measured molecules include proteins, peptides, metabolites, and other endogenous compounds.
Therefore, CEST imaging can add valuable information to the field of oncology, neurology, and musculoskeletal imaging \cite{Jones2018}.
Most notable are the detection \cite{Goerke2019}, grading \cite{Zhang2018}, mutation prediction \cite{Paech2018}, and monitoring of brain tumors \cite{Leung2024}, the assessment of cartilage degeneration \cite{Abrar2021}, and the detection of ischemic stroke via pH imaging \cite{Zhou2024}.

Although CEST MRI has been shown to be a versatile tool for many clinical applications, current state-of-the-art techniques are limited by three challenges:
First, conventional techniques acquire a series of images at different saturation frequencies that requires breaks in the measurement process (recovery time $T_{rec}$) between each offset to ensure full $T_1$ relaxation.
This markedly prolongs the acquisition times. 
Second, the technique is limited to short readouts to reduce the influence of relaxation effects.
This reduces the amount of data that can be acquired during one repetition as well as the signal-to-noise ratios (SNR) and spatial resolution for a fixed measurement time.   Third, most conventional CEST MRI techniques only provide qualitative image contrast, which combine a variety of different tissue and hardware characteristics. Influences include the used sequence, saturation parameters, relaxation rates of the tissue, and the chosen CEST metric,  which in current clinical practice is often the asymmetry analysis ($MTR_{asym}$) \cite{Zhou2022}. Asymmetry analysis is prone to confounding effects such as the spillover effect, and $T_1$ relaxation and results in a mixed contrast between the amide proton CEST effect and signal from nuclear Overhauser enhancement (NOE)\cite{Sun2024}.   Therefore, results from different studies, operators, and systems are difficult to compare, and various techniques have been proposed to improve the quantification of the CEST effect \cite{Kim2015}. These techniques range from post-processing methods such as Lorentzian fitting \cite{Zaiss2011} or advanced (inverse) CEST metrics \cite{Zaiss2014} , to different or multiple acquistions with additional evaluation schemes as in omega-plot analysis \cite{Wu2015} or Magnetic Resonance Fingerprinting (MRF) CEST \cite{Cohen2018}\cite{Cohen2023}. The aim of these techniques is to isolate the CEST effect from confounding factors, such as $T_1$, and provide a more specific and quantitative measure of the exchange processes, which can be more directly related to the underlying physiology and pathology.    

Zaiss et al. introduced an approach based on the inverse Z-spectrum, which is applicable to conventionally measured CEST data sets \cite{Zaiss2014}. This metric, called $MTR_{Rex}$, is defined as \begin{equation}
  MTR_{Rex} = \frac{1}{Z_{lab}} - \frac{1}{Z_{ref}}  = \frac{R_{ex}}{\cos^2{(\theta)}R_1}~,
\end{equation}
where $Z_{lab}$ is the label Z-spectrum and $Z_{ref}$ is the reference Z-spectrum.
This eliminates influence of the spillover effect, 
symmetric magnetization transfer (MT),
and $R_2$ relaxation.
Thus, the $MTR_{Rex}$ depends only on the exchange dependent relaxation rate $R_{ex}$,
the saturation ($\cos^2{(\theta)}$), and longitudinal relaxation rate $R_1=\frac{1}{T_1}$.  The influence of $R_1$ can be removed by multiplying $MTR_{Rex}$ with $R_1$ to obtain the apparent exchange-dependent relaxation \begin{equation}MTR_{AREX} = MTR_{REX}\cdot R_1 = \left(\frac{1}{Z_{lab}} - \frac{1}{Z_{ref}}\right)\cdot R_1 =  \frac{R_{ex}}{\cos^2{(\theta)}}~,\end{equation} which is the closest contrast to the theoretically expected $R_{ex}$ \cite{Zaiss2014}.
The derivation of $MTR_{AREX}$ requires the assumption of a steady-state Z-spectrum, which is not given in most CEST experiments in practice as it requires 
a long saturation time.
Therefore, $MTR_{AREX}$ is typically affected by model mismatches due to imperfect recovery between saturations \cite{Zaiss2014} \cite{Sun2024}.
The quasi steady-state correction (QUASS) offers an alternative, by estimating the steady-state spectrum from a given transient spectrum with 
known saturation parameters and $T_1$.
This enables a calculation of $MTR_{AREX}$ from transient spectra 
measured in a time constrained setting \cite{Sun2021} \cite{Sun2024},  
but requires prior knowledge of $T_1$.
Thus, a $T_1$ map has to be measured in a separate calibration scan increasing the measurement time and introducing possible errors due to misalignment between volumes and measurements.\\
In this work, we present a combined sequence and signal model for CEST, which 
addresses the challenges of conventional CEST measurements described above.
The overall acquisition time is shortened by rendering breaks in the measurement process between offsets obsolete. 
Relaxation effects during the measurement are taken into account using a time encoded FLASH readout after an inversion pulse.
Furthermore, the use of a stack-of-stars (SOS) readout allows for measurement of a 3D slab and therefore, greater coverage in a single acquisition.

In previous works CEST saturation was combined with an inversion pulse, but these methods added the inversion pulse before the saturation, thus changing the measured contrast \cite{Vinogradov2011} \cite{Jin2012}. 
The proposed model extends the approach by Chen et al. \cite{Chen2024} using water pre-saturation.
Instead of requiring to start from a vanished magnetization for each offset, the developed technique allows for starting the CEST saturation at the steady-state of a Look-Locker readout.
This transient Z-spectra is modeled using the equations given
in \cite{Zaiss2022} and the Look-Locker model \cite{Deichmann2005} for inversion recovery \cite{Roeloffs2016}.
The derived signal model is combined with model-based reconstruction, which allows for the incorporation of prior knowledge about the signal evolution and the underlying physics into the reconstruction process \cite{Wang2021}.
Model-based reconstruction is widely used in MRI for various applications \cite{Block2009}\cite{Sumpf2011}\cite{Roeloffs2016}\cite{Maier2019},
but to our knowledge has not yet been applied to CEST imaging.
Here, we use a model-based reconstruction approach for direct quantification of the exchange dependent relaxation rate $R_{ex}$ and the longitudinal relaxation rate $R_1$ from the measured data.

The proposed method was validated using a numerical phantom and compared to conventional CEST measurements and Look-Locker $T_1$-mapping in phantom and in vivo experiments on a clinical 3T scanner.

\section{Methods}
The developed technique consists of a sequence combining CEST saturation with a radial FLASH readout, a new analytical signal model derived from the Bloch-McConnell equations, and a model-based reconstruction. In the following, the individual parts are discussed in more detail.
\subsection{Proposed Sequence}
\label{sec:meth:seq}
The sequence consists of three blocks that are repeated for each of the acquired off-resonance frequency offsets of the z-spectrum. 
\begin{itemize}
	\item  First, off-resonant CEST saturation is applied using Optimal Control (OC) optimized pulses, comprising 15 pulses with a duration of = $t_p = 100$ ms, delay of $t_d=$1ms and $B_{1,rms}=1$ $\mu$T. This has been selected for its robust performance over $B_0$ and the close resemblance of the resulting spectrum to the desired continuous wave saturation, which is technically not feasible on a clinical scanner  \cite{Stilianu2024}. More details on the pulse design and optimization can be found in \cite{Stilianu2025}.
	\item  Second, a hyperbolic secant inversion pulse is applied with a duration of $8$ ms and pulse parameters  $T\Delta f = 10$, $\beta = 800$ rad/s, $\mu=4.9$ and $A_0 = 14 ~\mu$T following \cite{Bernstein2004}.
	\item Third, a golden-ratio based angle \cite{Winkelmann2007}, partition-aligned, 3D stack-of-stars FLASH readout is performed for encoding of the resulting Z-spectrum and $T_1$ information.
\end{itemize}

The inverted Z-spectrum is the starting point of each Look-Locker readout and, since no
recovery time ($T_{rec}$) between offsets is used to minimize the acquisition time,
the Look-Locker steady state ($M_{ss,LL}$ see Equation \eqref{eq:LL}) is the starting point for the CEST saturation of the next offset.

\begin{figure}[p]
  \centering
  \includegraphics[width=0.4\textwidth]{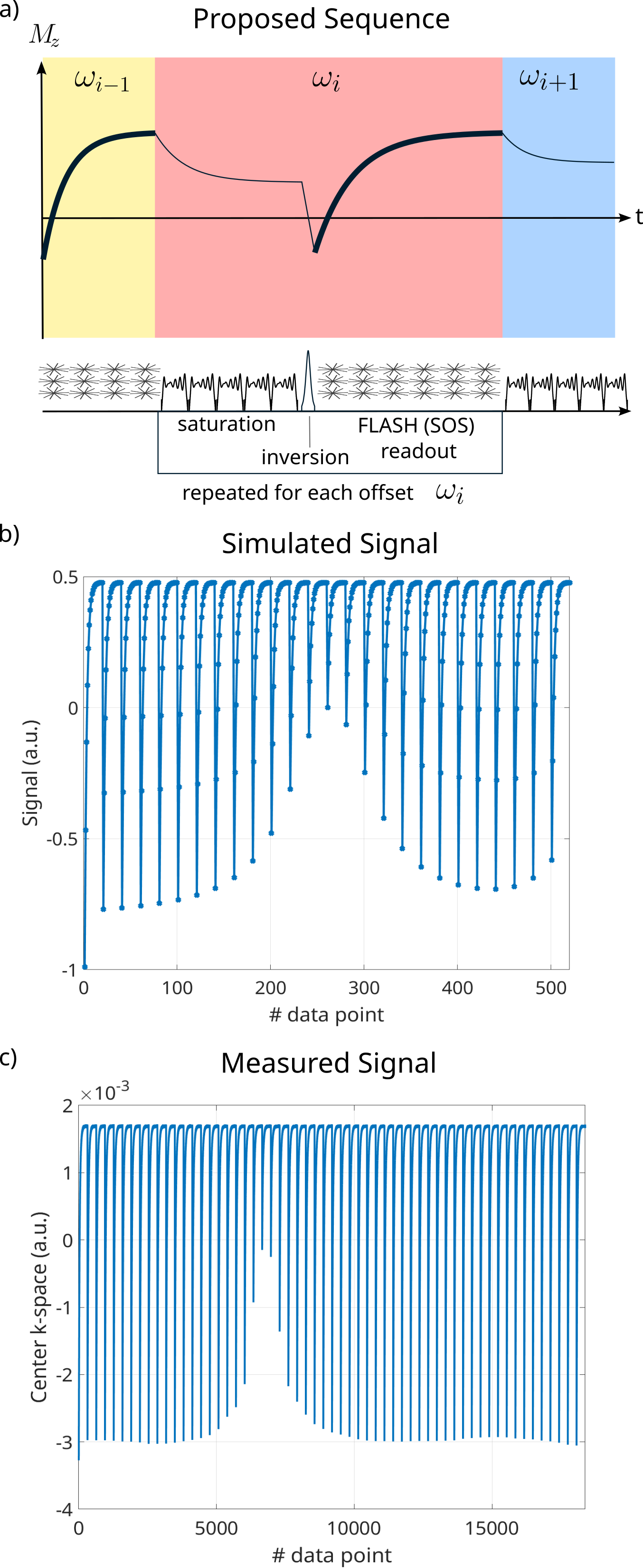}
  \caption{a) Proposed sequence, with saturation using the optimized OC pulses, the inversion pulse and the golden-ratio based stack-of-stars readout, repeated for each offset.
  Above the schematic magnetization evolution during those steps is shown. The bold lines indicate the magnetization captured by the readout.  b) Simulated signal evolution for a two pool model. (26 offsets from -4ppm to 4 ppm and 20 time points per offset) The inverted Z-spectrum can be seen as the starting point of the Look-Locker readout. c) Center k-space values (average of the two center points) for the phantom measurement.}\label{fig:res:00}
\end{figure}

Figure \ref{fig:res:00} shows an illustration of the three repeated phases of the sequence together with a schematic representation of the signal evolution and a simulated signal evolution as well as the measured signal for a phantom measurement.

\subsection{CEST Signal Model for Steady-State and non-Steady-State Experiments} 
The steady-state magnetization for a CEST experiment with offset $\omega$ can be derived from the Bloch-McConnell equations as
\begin{equation}
M_{ss,CEST}(\omega) = M_0P^2(\omega) \frac{R_{1w}}{R_{1\rho}(\omega)}~,
\end{equation}
when interpreted as an off-resonant spin-lock experiment as shown in \cite{Zaiss2022}. 
In this description $M_0$ is the equilibrium magnetization, $R_{1w}$ the observed longitudinal relaxation rate of water, $R_{1\rho}(\omega)$ the relaxation rate due to off-resonant saturation, as the magnetization is locked along the vector tiled from the z-axis by the angle $\theta$, which is given by 
$\cos(\theta) = \frac{\omega}{\sqrt{\omega^2+\omega_1^2}}$ and leads to the corresponding projection factor $P^2(\omega)$. $P^2$ is defined by $\frac{\omega^2}{\omega^2 + \omega_1^2}$, where $\omega_1 = \gamma B_1$, with the gyromagnetic ratio $\gamma$ and $B_1$ of the saturation pulse \cite{Zaiss2022}. 

The relaxation rate $R_{1\rho}(\omega)$ can be described as the sum of the water relaxation rates $R_{1w}$ and $R_{2w}$ when taking the projection angle $\theta$ and $R_{ex,i}(\omega)$, the influence of the  exchange processes on the relaxation for each of $N$ pools into account \cite{Zaiss2022}
\begin{equation}
R_{1\rho}(\omega) = R_{1w}cos^2(\theta) + R_{2w}cos^2(\theta) + \sum_{i=1}^{N} R_{ex,i}(\omega)~. 
\end{equation}

The exchange dependent relaxation rate $R_{ex}(\omega)$, for each of the N pools, can be modeled with a Lorentzian lineshape
\begin{equation}
R_{ex}(\omega) = a\frac{\Gamma^2/4}{\Gamma^2/4 + (\omega -  \delta \omega_i)}
\end{equation}
with amplitude 
\begin{equation}\label{eq:amplitude}
a = f_s k_s \frac{\omega_1^2}{k_s(k_s+R_{2s})+\omega_1^2}
\end{equation}
and linewidth
\begin{equation}
\Gamma = 2\sqrt{(R_{2s}+k_s)^2+\frac{R_{2s}+k_s}{k_s}\omega_1^2}~,
\end{equation}
where $f_s$ describes the relative fraction, $k_s$ the exchange rate, and $R_{2s}$ the relaxation rate of the exchanging pool  \cite{Zaiss2022}.
 
We define the ratio of the steady-state and equilibrium magnetization as the CEST steady-state fraction $f_{ss,CEST}(\omega)$
\begin{equation}
f_{ss,CEST}(\omega) =  \frac{M_{ss,CEST}(\omega)}{M_0} = P^2(\omega) \frac{R_{1w}}{R_{1\rho}(\omega)}~.
\label{eq:ss_frac}
\end{equation}

For a non-steady-state experiment, e.g. a transient sequence, we can formulate the signal model of the resulting spectrum as
\begin{equation}
M_{\text{trans},CEST}(\omega) =[M_{rec} - M_{ss,CEST}(\omega)]e^{-T_{prep}R_{1\rho}(\omega)}  + M_{ss,CEST}(\omega)
\end{equation}
with starting magnetization $M_{rec}$ and $T_{prep}$ the saturation time \cite{Zaiss2022}.
Note that depending on $T_1$ and the exchange parameters $T_{prep}$ is not long enough to reach full saturation. 

By using the left side of Equation \eqref{eq:ss_frac} and $f_{rec} = \frac{M_{rec}}{M_0}$ we end up with
\begin{equation}
    M_{\text{trans},CEST}(\omega) =M_0\left(\left[ f_{rec} - f_{ss,CEST}(\omega)\right]e^{-T_{prep}R_{1\rho}(\omega)}  + f_{ss,CEST}(\omega)\right)~.\label{eq:CEST_frac}
\end{equation}

\subsection{Combined CEST and Look-Locker Signal Model}

The Look-Locker signal equation for an inversion-recovery (IR) experiment with a continuous FLASH readout, with the assumption of perfect spoiling, reads as \cite{Look1970}\cite{Deichmann2005}
\begin{equation}
M_{\text{trans},LL}(t) = M_{ss,LL} - (M_0 + M_{ss,LL})e^{-tR_1^*}~. \label{eq:LL}
\end{equation}
Here, $R_1^*$ is the observed Look-Locker relaxation rate and $M_{ss,LL}$ is the steady-state magnetization, which is reached after sufficiently long pulse train. 
The starting point, for assumed perfect inversion, is $-M_0$ at $t=0$ s. $R_1^*$ describes combined effect of the longitudinal relaxation rate of water $R_1$ and the relaxation rate $R_1'$, which entails the influence of the readout. Here, $R_1' = 1/T_R \log(\cos(\alpha))$ with the flipangle $\alpha$ and  repetition time $T_R$ \cite{Deichmann2005}.
Therefore, the steady-state magnetization can be described by 
\begin{equation}
    M_{ss,LL} = M_0\frac{R_1}{R_1 + R_{1}'}~,
\end{equation}
which can be exploited to reformulate Equation \eqref{eq:LL}
\begin{equation}
    M_{\text{trans},LL}(t) = M_0 \left[f_{ss,LL} - (1 +f_{ss,LL})e^{-t (R_1 + R_1')} \right]\label{eq:LL_frac}
\end{equation}
with the steady-state fraction $f_{ss,LL} = M_{ss,LL}/M_0 = \frac{R_1}{R_1 + R_{1}'}$.
    
By replacing the starting magnetization of the Look-Locker experiment (the constant 1 in Equation \eqref{eq:LL_frac}) with the transient Z-spectrum from Equation \eqref{eq:CEST_frac}, we obtain a combined signal model for CEST saturation from incomplete recovery followed by a FLASH readout
\begin{equation}
    M(\omega,t) = M_0 \left[f_{ss,LL} - \left( \left[f_{rec}(\omega) - f_{ss,CEST}(\omega)\right]e^{-T_{prep}R_{1\rho}(\omega)}  + f_{ss,CEST}(\omega) +f_{ss,LL}\right)e^{-t (R_1 + R_1')} \right]~.\label{eq:CEST_transient}
\end{equation}

As mentioned before, the measurement is repeated for every offset.
Note that, except for the first measured offset, the starting magnetization for the CEST saturation is the Look-Locker steady-state 
from the previous offset
\begin{equation} 
    f_{rec}(\omega) =
  \begin{cases}
      1   & \quad \text{if } \omega^i \text{  is  } \omega^0\\
     f_{ss,LL}   & \quad \text{else}
  \end{cases}~.
\end{equation} 

\subsection{Model-Based Reconstruction}
The model-based reconstruction solves the following inverse problem \cite{Block2009}\cite{Sumpf2011}\cite{Roeloffs2016}
\begin{equation}\label{eq:generalProblem}
    \hat{u} = \argmin_u||\mathcal{P}\mathcal{F}\mathcal{C}\mathcal{M}(\omega,t,u)-d_k||^2_2 + \gamma_{R} TGV_j(u)~,
\end{equation}
where $d_{k}$ is the measured k-space data, $u$ is the parameter vector $[M_0,R_1, R_1', R_2, \omega_0, a_i,\Gamma_i]$, $\mathcal{M}(\omega,t,u)$ is the signal model from Equation \eqref{eq:CEST_transient}, $\mathcal{C}$ is the coil operator, $\mathcal{P}\mathcal{F}$ describes the non-uniform Fourier transform (nuFFT)\cite{Fessler2003}, and $\gamma_{R}$ is the weighting parameter  of the Total Generalized Variation (TGV) regularization \cite{Bredies2010}. 
TGV is an optimization problem itself, which is described by the following equation
\begin{equation}
TGV_j(u) = \min_v \alpha_0||\nabla u - v ||_1 +  \alpha_1|| \mathcal{E} v||_1~.
\end{equation}
Here, $\nabla$ is the finite forward difference, $\mathcal{E}$ the symmetrized  derivative $\mathcal{E} v = \frac12 (\nabla v + \nabla v^T)$, and $\alpha_0>0$ and $\alpha_1>0 $ are scalars balancing the first and second derivative, which are kept at a constant ratio of $\alpha_0/\alpha_1 = 1/2$ \cite{Knoll2011}. A more comprehensive description of the model-based reconstruction using TGV regularization can be found in \cite{Maier2019} and \cite{Huemer2024}.

\subsection{Data Acquisition - Numerical Tube Phantom}
\label{sec:meth:tube} 
A two-pool numerical phantom (CEST pool at 4.2 ppm) was implemented with eleven different water $T_1$ and CEST concentration combinations. Water $T_1$ values were chosen to be 500, 600, 700, 800, 900, 1000, 1100, 1200, 1300 and 1500 ms. 
The CEST concentrations were chosen to be 0, 0.1, 0.2, 0.3, 0.4, 0.5, 0.6, 0.6, 0.7, 0.8 and 0.9 $\%$ relative fraction. 
The exchange rate was set to 200 Hz, $T_2$ was set to 110 ms and the $T_1$ value of the exchange protons was set to 1000 ms.
The magnetization evolutions for these parameters were simulated using a Bloch-McConnell simulation \cite{Graf2021} implemented in MATLAB (2022b, MathWorks) that has been validated in a previous study \cite{Schuenke2023}.
The image-space phantom was constructed by using the geometry of a tube phantom implemented in the Berkeley Advanced Reconstruction Toolbox (BART) \cite{BART} and setting each pixel to the corresponding signal evolution. The resulting image-space phantom was used to test the noise-free image-space fit of the proposed signal model.  

Using the same geometry and signal the phantom was also constructed in k-space twice. Once with a fully sampled Cartesian sampling pattern and once with a non-Cartesian sampling pattern consisting of three -- by a golden-ratio based angle rotated -- radial spokes per image. 
Gaussian noise with variance of 5 $\%$ of the maximum signal intensity was added to the k-space data in the real and imaginary part. The fully sampled Cartesian data was then reconstructed using a standard Fourier transform and coil combined using the ground truth coil sensitivities. The Cartesian dataset was used to evaluate the image-space fit of the proposed signal model with added noise and the non-Cartesian dataset was used to evaluate the model-based reconstruction.  

Furthermore, a reference data-set was constructed by simulating the spectra with the same saturation parameters, but with a recovery time $T_{rec}$ of $3.5$ s. 
Again, each pixel of the phantom was set to the corresponding spectra, to construct the image-space phantom.  Again the same phantom was also constructed in k-space using a fully sampled Cartesian sampling pattern. Noise with variance of 5 $\%$ of the maximum signal intensity was added to the k-space data in the real and imaginary part and the data was reconstructed using a standard Fourier transform and coil combined using the ground truth coil sensitivities. In this way, reference data-sets with and without noise were constructed. QUASS correction was then applied according to \cite{Sun2021} to both reference data-sets. The QUASS corrected data was further evaluated using a pixel-wise two-pool Lorentzian fit, using the lsqcurvefit function in MATLAB implementing the Levenberg-Marquardt algorithm.
From the fit results $MTR_{AREX}$ was calculated at $\omega_i=4.2$ ppm using 
\begin{equation}
  MTR_{AREX}(\omega_i) = \left(\frac{1}{Z_{lab}(\omega_i)} - \frac{1}{Z_{ref}(\omega_i)}\right)R_1~,
  \label{eq:MTRAREX} 
\end{equation}
where $Z_{lab}$ is the spectrum calculated from the Lorentzian fit and $Z_{ref}$ is the reference spectrum calculated from the Lorentzian fit, but with the amplitude of the CEST pool set to 0. 
For $R_1$ the ground truth $R_1$ map from the simulation input was used.

\subsection{Data Acquisition - Phantom and in vivo Measurements}

A phantom was constructed with 50 ml Falcon tubes filled with different concentrations of Iohexol  (GE Healthcare, Chicago, United States of America).
This substance shows a CEST effect at 4.2 ppm and the tubes were additionally doped with Gadovist (Bayer Vital GmbH, Leverkusen, Germany) to produce the desired range of $T_1$ values.

The proposed sequence was implemented in MATLAB as described in Section \ref{sec:meth:seq} using the open-source pulse programming standard Pulseq \cite{Layton2017}\cite{Herz2021}.

For the phantom measurement the proposed sequence was used with the saturation as described in Section \ref{sec:meth:seq} for 58 offsets and the following readout parameters for the 3D SOS FLASH readout: FOV 192x192x20 mm$^3$, base resolution = 128, readout oversampling factor = 2, $T_E= 2.1$ ms, $T_R=4.2$ ms, $\alpha=6$°, phase spoiling with quadratic phase increment of 84° \cite{Weinmller2025}, slab thickness=18 mm, 3 partitions, golden-ratio based spoke increment (111.246°) with aligned partitions and a total acquisition time of 5:30 min.

A conventional CEST measurement was performed with identical saturation and offsets, but differing $T_{rec}$ of 3.5 s. The readout consisted of a centric reordered 3D GRE with: FOV of 192x192x20mm$^3$, base resolution = 128x128x3, $\alpha=4$°, $T_E= 2.56$ ms, $T_R=5.1$ ms, slab thickness=18 mm and a total acquisition time of 6:51 min.

Furthermore, a WASABI \cite{Schuenke2017} sequence was measured for determining reference $B_1$ and $B_0$ maps using the same readout parameters as the conventional CEST sequences. 

For $T_1$ mapping, an inversion-recovery Look-Locker (IR-LL) SOS sequence was acquired using the same inversion pulse and readout parameters as the proposed sequence \cite{Roeloffs2016}.

The in vivo measurements were performed on a healthy volunteer using the same protocol as the phantom measurements including the proposed method, the conventional CEST measurement, the WASABI $B_1$ and $B_0$ reference and the IR-LL sequence.
The protocol only differed in the use of a larger FOV of 256x256x20mm$^3$ and a wider range of offsets from -300ppm to 300ppm instead of the previously used range from -5ppm to 6ppm.  
All experiments were performed on a clinical 3T MRI scanner (MAGNETOM Vida, Siemens Healthineers GmbH, Erlangen, Germany) using a 20-channel head-coil. 
Informed consent was obtained, and the study was approved by the local ethics committee.

\subsection{Data Evaluation - Model-based reconstruction and evaluation of parameter maps} 

The proposed signal model from Equation \eqref{eq:CEST_transient} was implemented in PyQMRI, a python-based open-source reconstruction toolbox \cite{Maier2020}.
PyQMRI implements an iteratively regularized Gauss-Newton (IRGN) method using primal-dual splitting to solve the non-linear inverse problem given in Equation \eqref{eq:generalProblem}.

For the numerical phantom, the parameter maps were evaluated against the ground-truth relative fraction of the CEST agent. The fitted relative fraction can be calculated from the amplitude $a_{fitted}$ of the fitted $R_{ex}$ by rewriting Equation \eqref{eq:amplitude} as
\begin{equation}
  f_{s,fitted}  =  \frac{a_{fitted}}{k_s \frac{\omega_1^2}{k_s(k_s+R_{2s})+\omega_1^2}}~.
  \label{eq:fittedconc}
\end{equation}

For the measurements, the reference $T_1$ maps were calculated from the IR-LL sequence with a model-based reconstruction based on the Look-Locker signal model from Equation \eqref{eq:LL_frac} also implemented in PyQMRI.

The fully sampled conventional CEST measurements were reconstructed using the inverse Fourier-transform, coil-combined with the root-sum-of-square method, and denoised using PCA \cite{Mennecke2022}, while keeping the first 15 principal components. For the in vivo dataset, motion correction was performed with BART, registering the images using affine transformation (function affinereg) and cubic interpolation (function interpolate).  Afterwards, the QUASS corrected spectra were calculated using the reference $T_1$ maps and a pixel-wise multi-pool Lorentzian fit was applied using 2-pool (water and CEST at 4.2 ppm) and 4-pool models (water, CEST at 3.5 ppm, NOE at -3.5 ppm and MT at -2 ppm), for the phantom and in vivo measurements, respectively. 
$MTR_{AREX}$ was determined using the Lorentzian fit results, as described in Equation \eqref{eq:MTRAREX}, at 4.2 ppm for the phantom and at 3 ppm for the in vivo measurements \cite{Liu2023}. All pre- and post-processing steps for the conventional measurements were implemented in MATLAB.

For the measured radial data, gradient delay correction \cite{Rosenzweig2019} was applied on the trajectory and SVD based coil compression \cite{Buehrer2007}\cite{Huang2008} to eight virtual channels was performed with BART \cite{BART}.  After applying the IFFT in partition direction, 80 spokes were selected for the reconstruction per partition and offset. The selected time points were chosen using the MATLAB function logspace in order to distribute more points to the start of the readout.  Using 80 spokes per offset results in a nominal undersampling factor of 2.5.   The reduced data-set was then reconstructed using the model implemented in PyQMRI \cite{Maier2020}.   PyQMRI solves the inverse problem given in Equation \eqref{eq:generalProblem} using an iteratively regularized Gauss-Newton (IRGN) method with primal-dual splitting.  Starting values for the regularization parameter $\gamma_{R}$ ($\gamma_{Start}$), minimal values ($\gamma_{min}$), and the reduction factor ($a$), which is applied after each Gauss-Newton iteration ($\gamma_i = a\times \gamma_{i-1}$, where $i$ is the Gauss-Newton step), can be found in supplemental Table S1. Also noted there are the number of Gauss-Newton steps performed for each reconstruction. For each of those steps a maximum of 200 primal-dual iterations were performed  

The resulting $T_1$ values of the proposed method were evaluated against the reference from the IR-LL sequence, while the CEST $R_{ex}$ amplitude was compared to the QUASS corrected $MTR_{AREX}$ of the conventional CEST measurement.

For all results, mean as well as standard deviation values were calculated for different region of interests (ROIs), and Bland-Altman plots were created to evaluate the agreement between the ground truth or reference and the fitted values of the proposed method.

\section{Results}
\subsection{Numerical Phantom}
\label{sec:res:tube} 
\begin{figure}[p]
    \centering
    \includegraphics[width=0.5\textwidth]{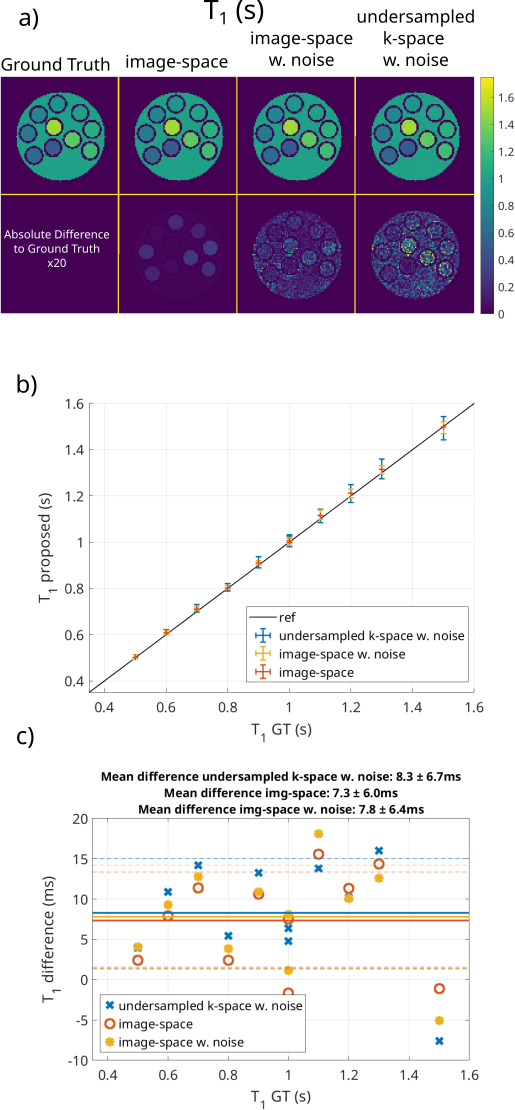}
    \caption{$T_1$ results for the proposed method for the numerical phantom. 
    a) shows the ground truth and the fitted parameter maps for the image-space fit with and without added noise and the model-based reconstruction from k-space.  
    In b) the mean and standard deviation for each region is displayed and c) shows a Bland-Altman-plot compared to the ground truth.}\label{fig:res:01}
  \end{figure}

Figure \ref{fig:res:01} shows the results for $T_1$ for the numerical phantom. 
The ground truth and the fitted parameter maps for the image-space fit, with and without noise, and the model-based reconstruction from k-space (with added noise) are displayed in Figure \ref{fig:res:01} a). 
There is no visible difference between the noise-free image-space results and the ground truth map. While the image-space fit with noise and the k-space reconstruction show a slightly lower SNR. The absolute difference between the ground truth and the fitted $T_1$ values is shown in the second row. Here, scaled by a factor of 20, the small deviations are visible. The added noise is also clearly visible for both the image-space and k-space fit. The mean and standard deviation for each region (individual tubes and background), compared in Figure \ref{fig:res:01} b), and the Bland-Altman plot shown in Figure \ref{fig:res:01} c), reveal close agreement between the ground truth and the fitted $T_1$ values. 
The mean deviation to the ground truth is 7.3 ms with a standard deviation of 6.0 ms for the image-space fit, 7.8 ms with standard deviation 6.4ms for the image space fit with added noise and 8.3 ms with a standard deviation of 6.7 ms for the model-based reconstruction from k-space.

\begin{figure}[p]
    \centering
    \includegraphics[width=0.65\textwidth]{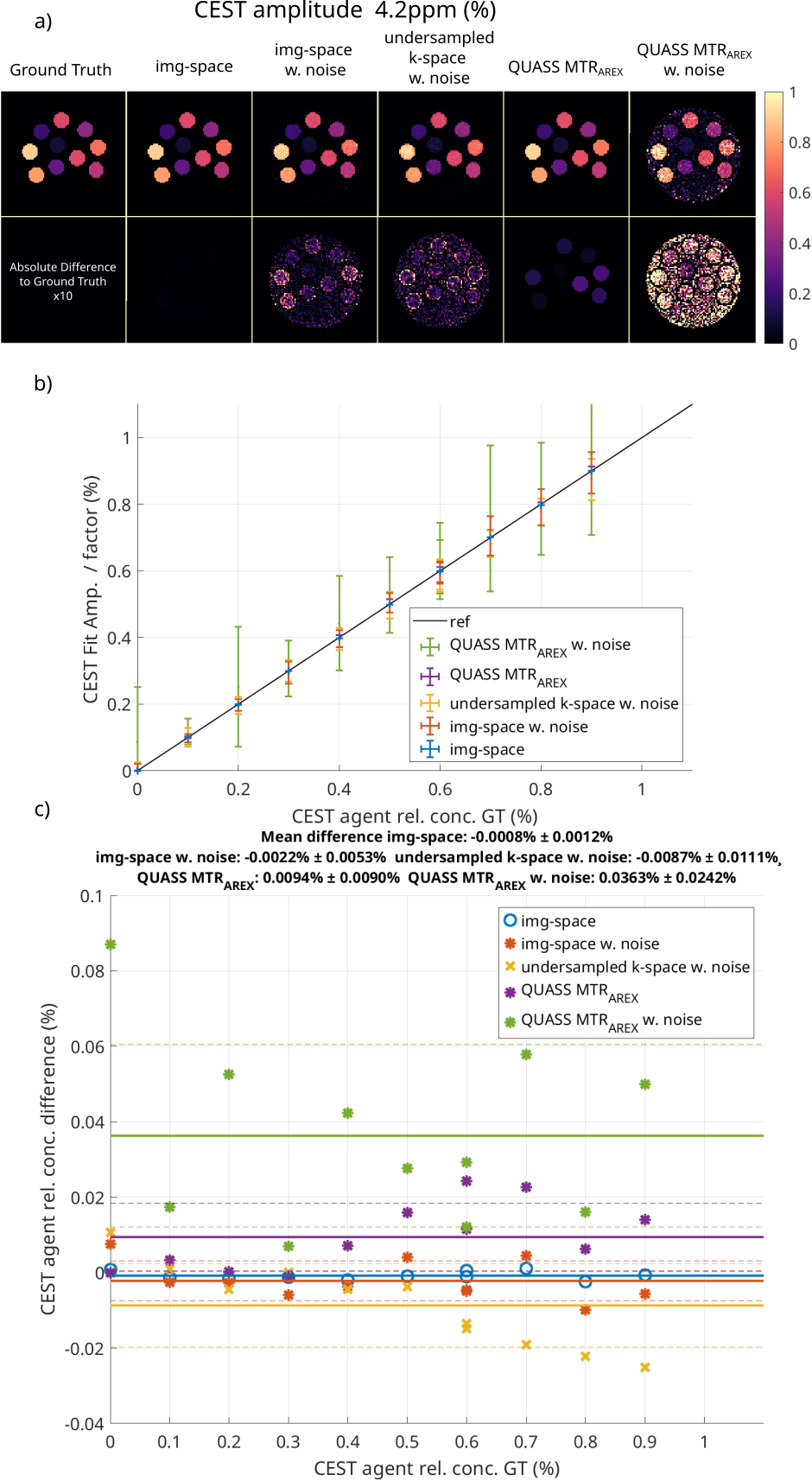}
    \caption{Relative CEST agent concentration results for the proposed method and calculated from $MTR_{AREX}$ for the numerical phantom. 
    a) shows the ground truth map, the image-space fit with and without noise, the parameter map obtained from the model-based reconstruction of k-space data and the calculated $MTR_{AREX}$ with and without noise. 
    In b) the mean and standard deviation for all regions is plotted against the ground truth and c) shows the Bland-Altman-plot for all three methods against the ground truth.}\label{fig:res:02}
\end{figure}

In Figure \ref{fig:res:02} a) the results for the relative CEST agent concentration (Equation \eqref{eq:fittedconc}) are shown for the image-space fit, with and without added noise, and k-space model-based reconstruction with the proposed method.
The figure also presents the reference QUASS corrected $MTR_{AREX}$ once with added noise and once without noise  The absolute difference between the ground truth and the fitted relative CEST agent concentration is shown in the second row of Figure \ref{fig:res:02} a. Even though the difference is scaled by a factor of 10, no difference is visible for the noise free image-space fit, while added noise is clearly visible for both the image-space fit with noise and the k-space fit. The QUASS corrected $MTR_{AREX}$ shows some systematic deviation and with added noise clear deviations from the ground truth appear. The mean and standard deviation for each region, plotted in Figure \ref{fig:res:02} b), and the Bland-Altman plot added in Figure \ref{fig:res:02} c), show small differences against the ground truth.
The image-space fit shows a mean deviation in relative CEST agent concentration, when rounded to the fourth significant digit, of -0.0008 $\%$pt with a standard deviation of 0.0012 $\%$pt.  For the image-space fit with added noise the mean deviation is -0.0022 $\%$pt with a standard deviation of 0.0053 $\%$pt. The model-based reconstruction from k-space shows a negative bias with a mean deviation of -0.0087 $\%$pt and a standard deviation of 0.0111 $\%$pt. A slight negative trend for increasing CEST agent concentration is visible.   
The QUASS corrected $MTR_{AREX}$ shows a mean deviation of 0.0094 $\%$pt with a standard deviation of 0.0090 $\%$pt.  With added noise this deviation increases to a mean of 0.0363 $\%$pt and a standard deviation of 0.0242 $\%$pt.  \subsection{Phantom Measurements}
\label{sec:res:phan}

\begin{figure}[p]
    \centering
    \includegraphics[width=0.5\textwidth]{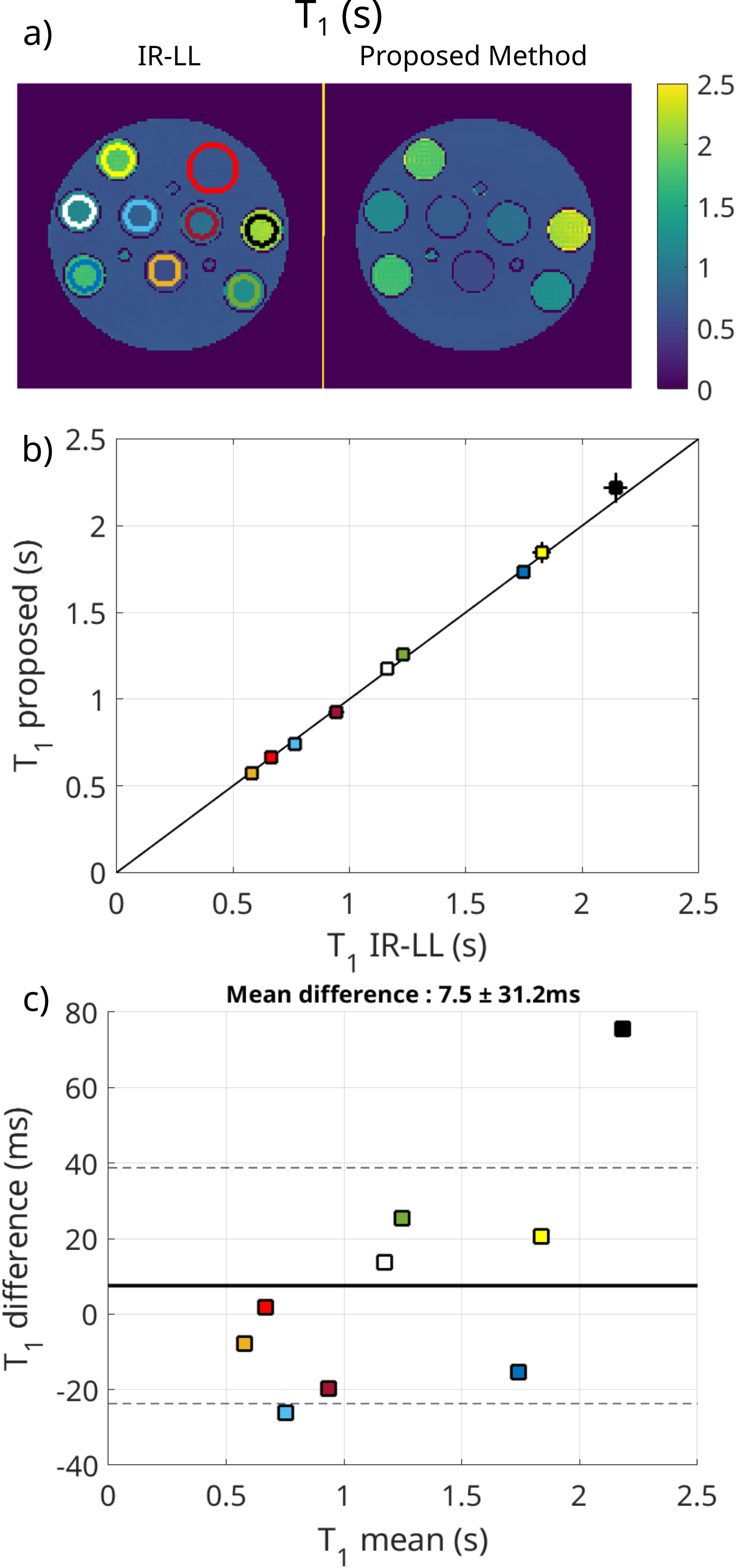}
    \caption{$T_1$ results for the phantom measurements. Comparison of the reference IR-LL sequence to the proposed method. 
    a) shows the parameter maps, in b) the mean and standard deviation are plotted and in c) a Bland-Altman plot comparing the two methods is displayed.}\label{fig:res:03}
  \end{figure}

Figure \ref{fig:res:03} shows and compares the measured $T_1$ maps of the phantom experiment determined with the proposed and reference method. ROIs are defined and colored for each of the tubes and the background. 
The maps in Figure \ref{fig:res:03} a) show no visible difference between both methods. 
The mean and standard deviation for each ROI is plotted in Figure \ref{fig:res:03} b) and in the Bland-Altman plot in Figure \ref{fig:res:03} c) comparing the two methods.
The standard deviation for both methods increases for ROIs with higher $T_1$ values.
Overall, the mean difference over all ROIs between the methods is 7.5 ms and the standard deviation is 31.2 ms.

\begin{figure}[p]
    \centering
    \includegraphics[width=0.5\textwidth]{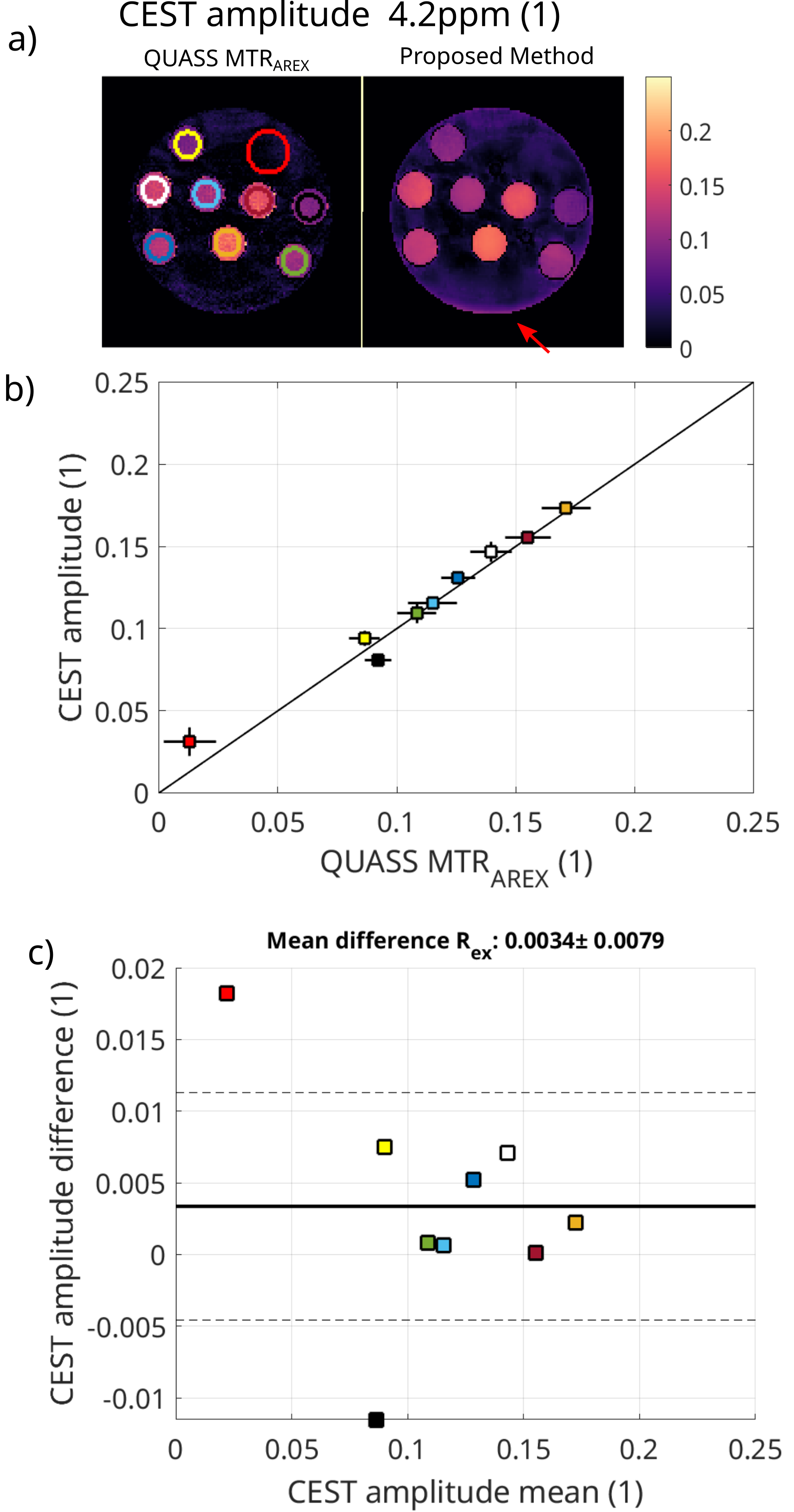}
    \caption{Comparison of the CEST amplitude results for the conventional CEST measurement and the proposed method. 
    a) shows the parameter maps for the QUASS corrected conventional method and the proposed method. The red arrow indicates higher erroneous values in the proposed method, likely caused by low coils sensitivity add the open end of the head coil.
    In b) the mean and standard deviation are shown using the conventional method as reference and in c) a Bland-Altman plot comparing the two methods is displayed. }\label{fig:res:04}
  \end{figure}

In Figure \ref{fig:res:04}, the comparison of the CEST amplitude results for the conventional CEST measurement and the proposed method is shown. 
The maps in Figure \ref{fig:res:04} a) show slight differences between the QUASS corrected conventional method and the proposed method. 
The conventional method shows higher noise, while the proposed method has a higher background signal and (towards the lower edge, marked by the red arrow) and increased homogeneity in the tubes.
This can be observed as well in Figure \ref{fig:res:04} b) showing an elevated mean value of the proposed method in the ROI with the lowest amplitude and increased standard deviation for the conventional method.
The Bland-Altman plot in Figure \ref{fig:res:04} c) shows a mean difference of 0.0034 with a standard deviation of 0.0079 for the proposed method compared to the conventional method for all ROIs.
The outliers represent the background and the highest $T_1$ value tubes.

\begin{figure}[p]
    \centering
    \includegraphics[width=1\textwidth]{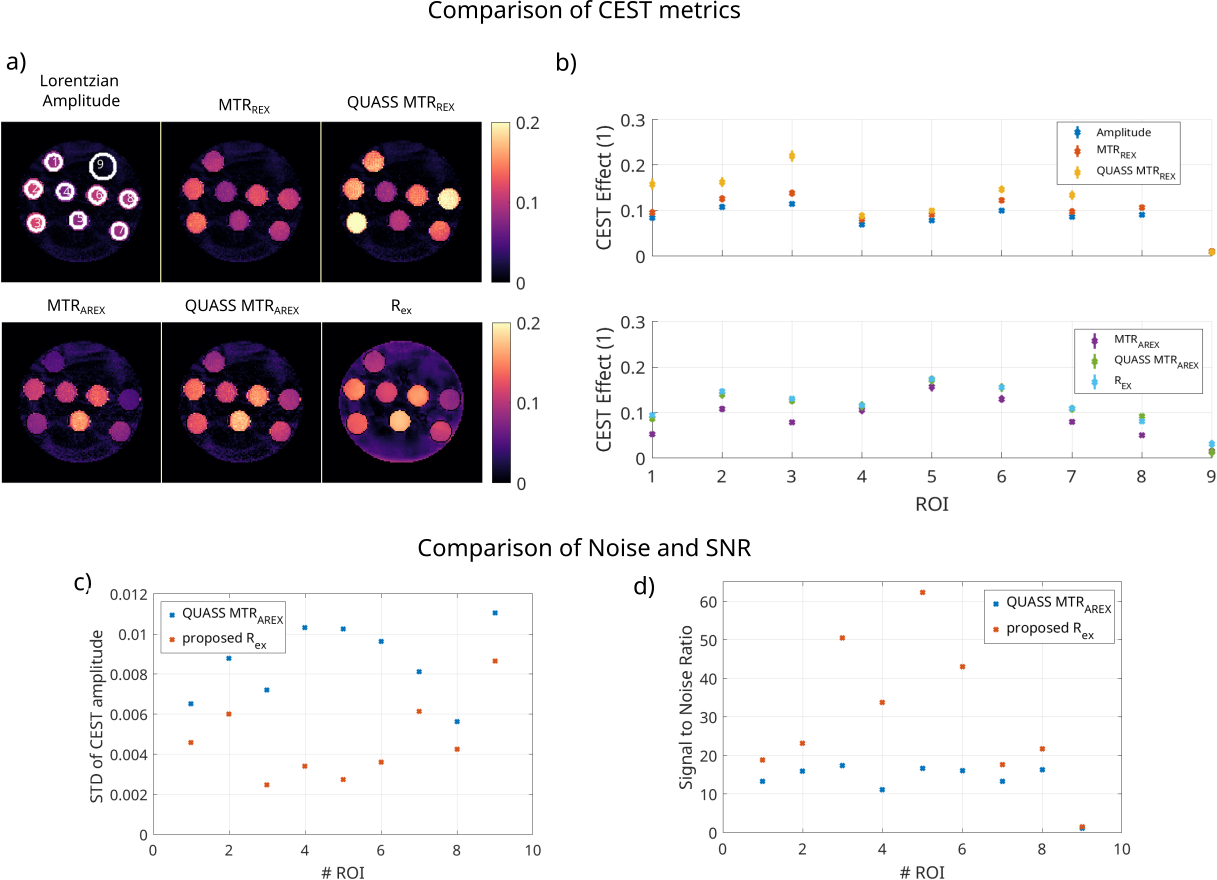}
    \caption{Comparison of the CEST metrics: For the conventional CEST measurement the Lorentzian fit amplitude, $MTR_{REX}$, QUASS corrected $MTR_{REX}$,  $MTR_{AREX}$ and QUASS corrected $MTR_{AREX}$ are shown and compared to the proposed method.
    a) shows the maps and b) the mean and standard deviation for each tube and the background. c) shows the standard deviation for each ROI for QUASS $MTR_{AREX}$ and the proposed method and d) compares the signal-to-noise ratio for the two methods.}\label{fig:res:05}
  \end{figure}

Evaluations of various further CEST metrics are presented in Figure \ref{fig:res:05}.
The Lorentzian fit amplitude, $MTR_{REX}$, QUASS corrected $MTR_{REX}$,  $MTR_{AREX}$, and QUASS corrected $MTR_{AREX}$ are compared to the proposed $R_{ex}$  method. 
Here, the deviations between possible metrics are apparent. 
Notable differences are present due to QUASS correction of the saturation time ($MTR_{REX}$ to QUASS $MTR_{REX}$ and $MTR_{AREX}$ to QUASS $MTR_{AREX}$) and the $T_1$ correction of the  $MTR_{AREX}$ ($MTR_{REX}$ to $MTR_{AREX}$ and QUASS $MTR_{REX}$ to QUASS $MTR_{AREX}$).
The proposed method $R_{ex}$ shows good agreement with QUASS $MTR_{AREX}$, which corrects for saturation time and $T_1$. In Figure \ref{fig:res:05} c) the standard deviation for each ROI is plotted, showing the lower standard deviation for the proposed method, while Figure \ref{fig:res:05} d.) shows the signal to noise ratio (SNR) for each tube. Again the proposed method shows higher SNR for all tubes.  The conventional CEST measurement combined with the IR-LL sequence took 6:30 min, while the proposed method took  5:30 min for the combined acquisition of $T_1$ and CEST $R_{ex}$. 

\subsection{In Vivo Measurements}
\label{sec:res:invivo} 

\begin{figure}[p]
    \centering
    \includegraphics[width=0.5\textwidth]{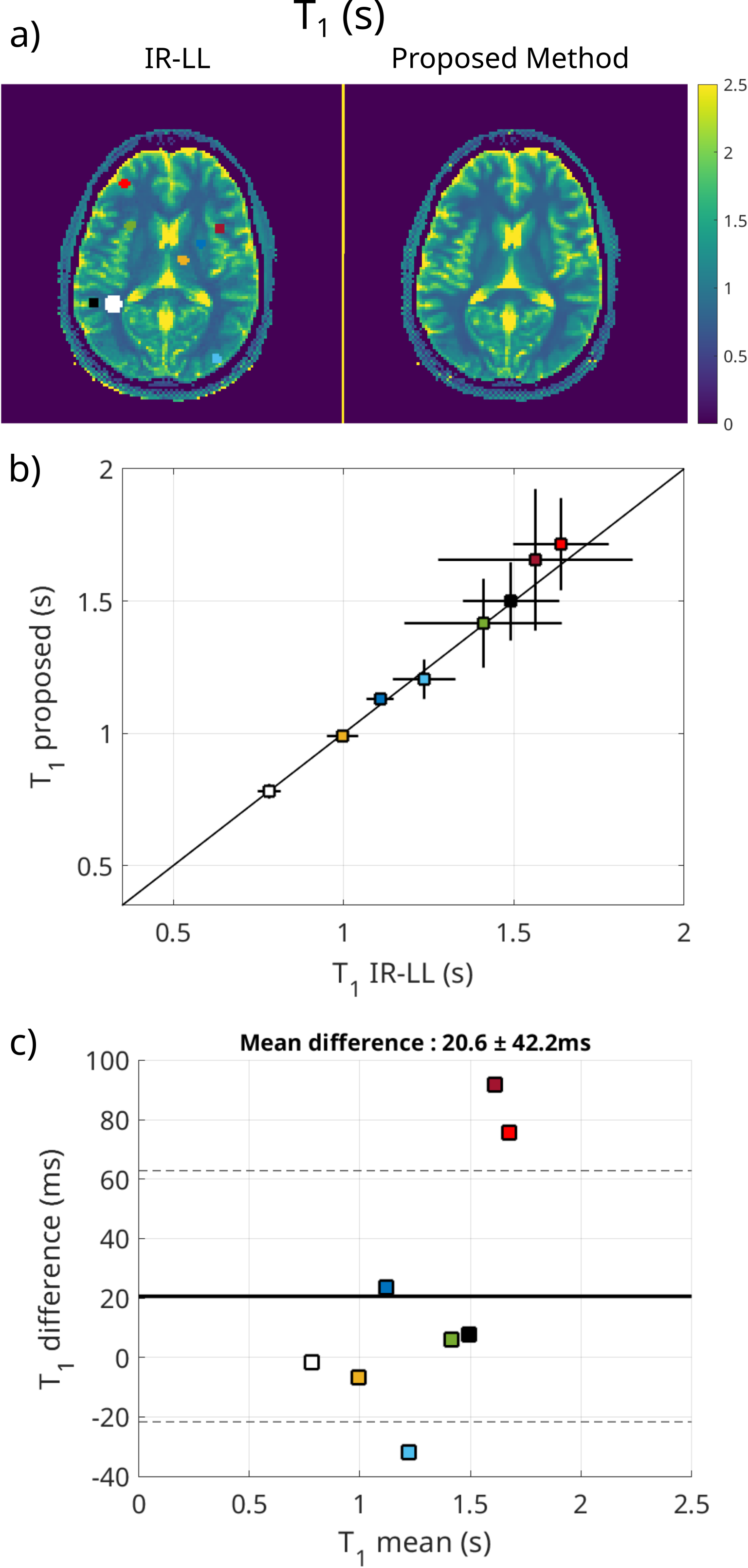}
    \caption{Comparison of the fitted $T_1$ values for the in vivo measurement. 
    a) shows the resulting parameter maps for IR-LL and the proposed method. In the IR-LL map the evaluated ROIs are marked in black. 
    b) displays the mean and standard deviations calculated from the ROIs and in c) the Bland-Altman plot comparing the methods is displayed.}\label{fig:res:06}
  \end{figure}

Figure \ref{fig:res:06} shows the results for $T_1$ for the in vivo measurement. In Subfigure \ref{fig:res:06}.a the maps of the reference IR-LL sequence and proposed method are shown. No apparent difference is visible.
ROIs for in detail analysis are defined and marked in the IR-LL map.
In Figure \ref{fig:res:06} b) both methods are compared by displaying the mean and standard deviation for each ROI. For both methods the standard deviation increases for ROIs with higher $T_1$ values.
The Bland-Altman plot in Figure \ref{fig:res:06} c) shows a mean difference over all ROIs between the methods of 20.6ms and a standard deviation of 42.2ms.

\begin{figure}[p]
    \centering
    \includegraphics[width=0.5\textwidth]{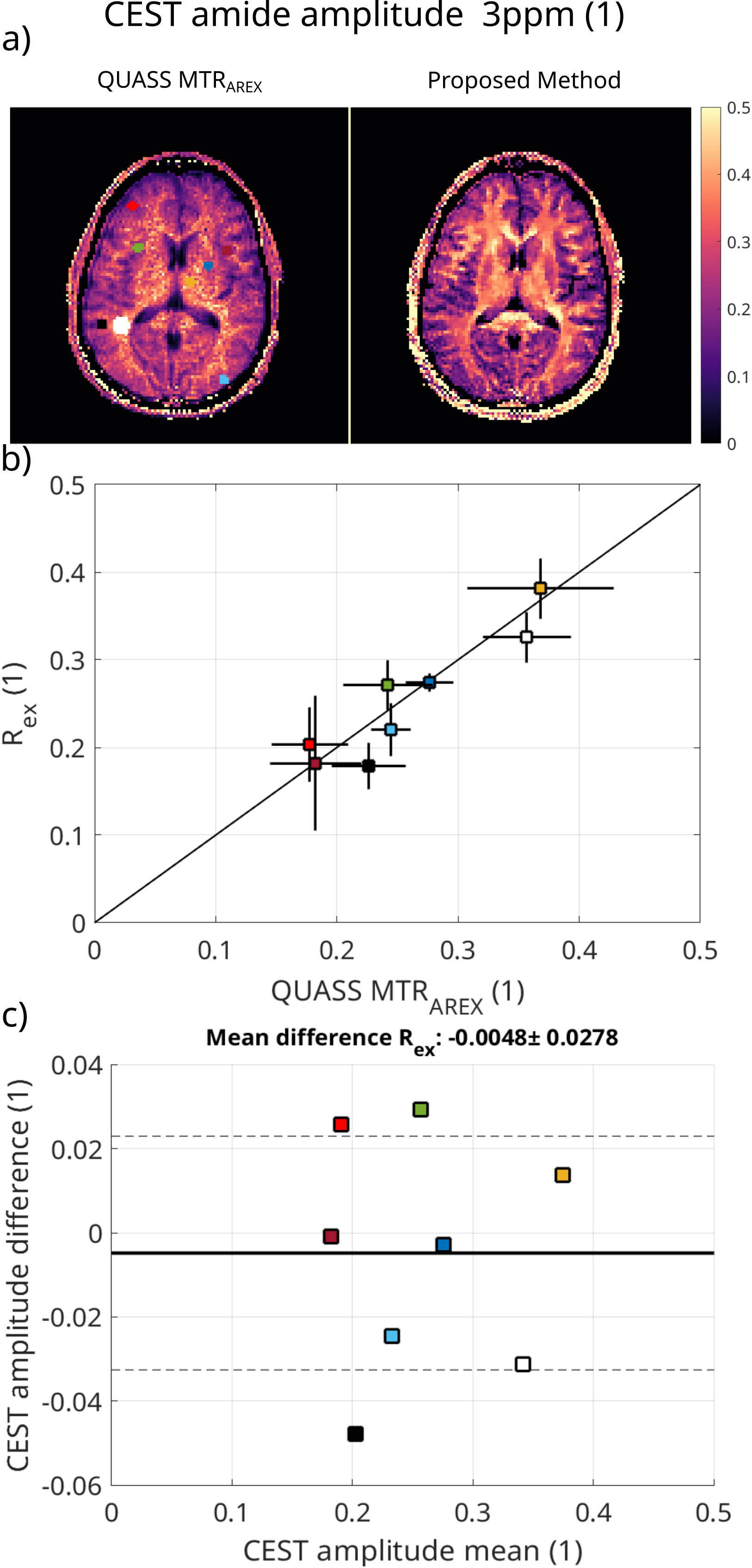}
    \caption{CEST amplitude results for the in vivo measurements. 
    a) shows the calculated QUASS $MTR_{AREX}$ map from the conventional method and the $R_{ex}$ map reconstructed from the proposed method. 
    In b) mean and standard deviation in the ROIs are displayed and in c) the Bland-Altman plot is shown.}\label{fig:res:07}
  \end{figure}

In Figure \ref{fig:res:07}, the comparison of the QUASS $MTR_{AREX}$ results for the conventional CEST measurement and the proposed method (amplitude of $R_{ex}$) is shown. The same ROIs presented in the $T_1$ maps are analyzed.
The resulting maps in \ref{fig:res:07}.a are in good agreement between both methods.
The conventional method has increased noise, while the proposed one provides a more homogenous signal.
Central regions of the brain show higher values for the proposed method, while some parts of gray matter have lower values. 
The proposed method shows sharper defined edges between tissue types as visible between deep gray matter and white matter.
The ROI analysis in \ref{fig:res:07}.b confirms a higher standard deviation of the conventional method and lower mean values with the proposed method. 
Mean difference over all ROIs between the methods is -0.0048 with a standard deviation of 0.0278, as shown in the Bland-Altman plot in Figure \ref{fig:res:07} c.).

\begin{figure}[p]
    \centering
    \includegraphics[width=0.8\textwidth]{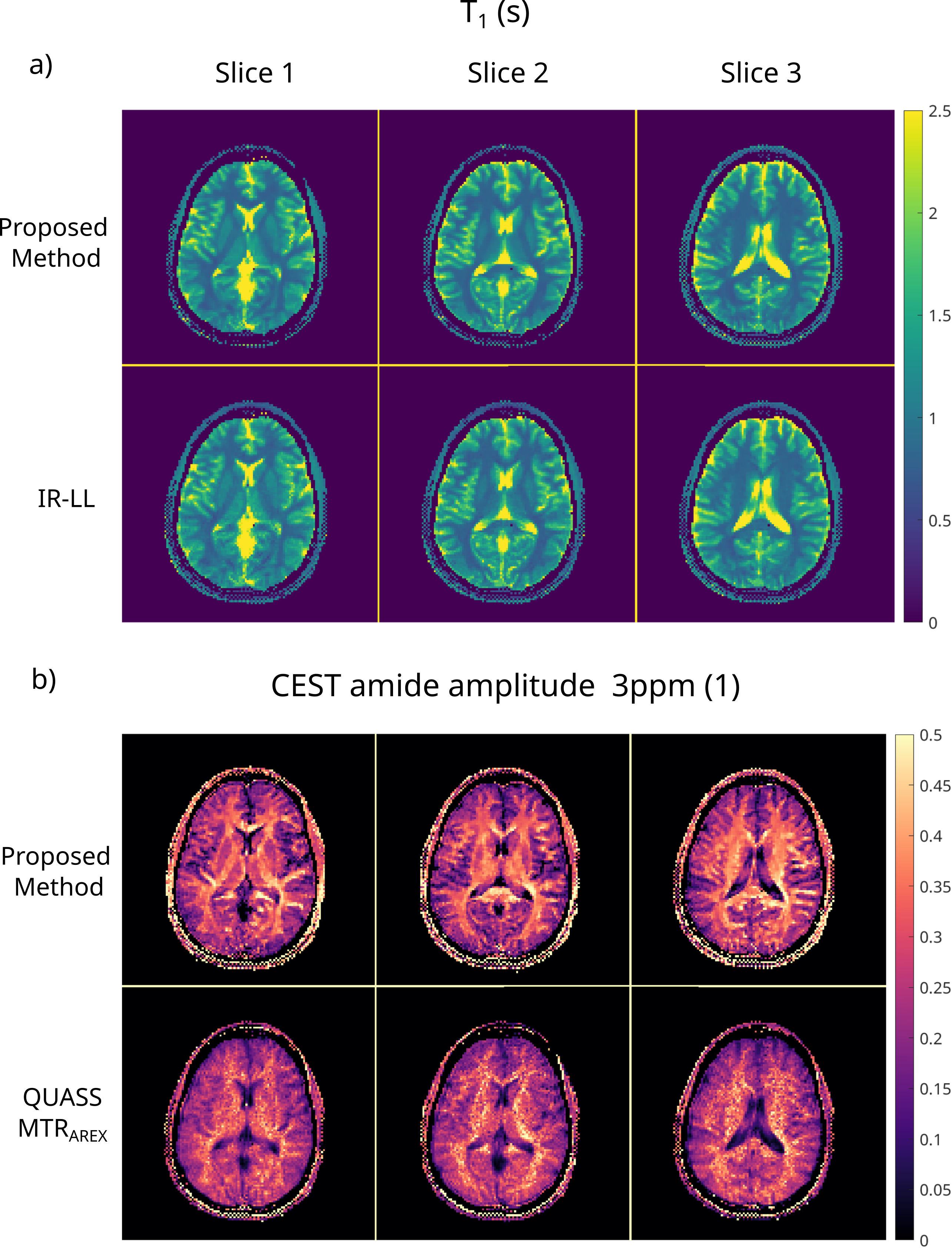}
    \caption{Parameter maps for all acquired slices for $T_1$ (a.), comparing IR-LL with the proposed method, and the CEST amplitude at 3ppm (b.), comparing the QUASS $MTR_{AREX}$ of the conventional measurement with the $R_{ex}$ of the proposed method.}\label{fig:res:08}
  \end{figure}

Maps with $T_1$ values and CEST amplitudes of all measured slices are presented in Figure \ref{fig:res:08}.a and (\ref{fig:res:08}.b, respectively. 
$T_1$ values are consistent over all slices with no observed difference between the two methods.
The CEST amplitude maps show good agreement between the two methods with increased noise for the conventional and increased homogeneity for the proposed method. The QUASS $MTR_{AREX}$ shows slightly higher values for the second and third slice, when being determined with the conventional technique.
Similar deviations between slices are not present with the proposed method.

The total acquisition time for the in vivo measurements was 7:10 min for the conventional CEST measurement combined with the IR-LL sequence, while the proposed method took 5:24 min. 

\section{Discussion}

\subsection{Numerical Tube Phantom}
For $T_1$ measurements we achieve a high similarity between the ground truth references for the image-space fit with and without noise and the model-based reconstruction from k-space. 
The deviations are expected to result from numerical errors in the fitting process and in the Bloch-McConnell simulation, used to generate the phantom.
The additional noise does not seem to influence the mean of the $T_1$ values, as the overall bias changes only slightly. The standard deviations are increased with higher $T_1$ values as expected.

The CEST relative concentration, shows values very close to the ground truth for the image-space fit without noise, resulting in no significant bias or variance.  Additional noise leads to an increase in the variance, as expected. The bias only increases slightly, from -0.0008 $\%$pt to -0.0022 $\%$pt, indicating that the image-space fit is robust to noise.      The slight negative trend for increasing CEST agent concentration for model-based reconstruction is probably due to the stronger regularization of the highly under-sampled data in this simulation (tree pokes per image). Strong regularization may reduce the step height for sudden changes, as represented within the numerical phantom.

The QUASS $MTR_{AREX}$ has positive bias and higher variance, which is probably caused by multiple numerical errors and their propagation, in the QUASS correction, multiple Lorentzian fitting, and the $MTR_{AREX}$ calculation, as no trend  can be observed. These steps also seem to be very sensitive to the added noise, as the variance increases and significant positive bias is introduced, leading to an overestimation of the CEST concentration. For all other methods, the maximum relative error (for non zero concentrations) is still under $5\%$. Here, the direct quantification of $R_{ex}$ is superior, as it incorporates the knowledge of the saturation time to estimate the steady-state CEST spectrum and involves only a single step that reduces the influence of numerical errors and noise propagation. This leads to a more robust and accurate quantification of the CEST agent concentration, with similar deviations even tough the numerical phantom has added noise and therefore significantly lower SNR.

\subsection{Phantom Measurements}
The phantom experiment shows similar results to the numerical phantom.
$T_1$ values show good agreement between the both methods over the whole range of $T_1$ values of the phantom. 
For the CEST maps, the background without any CEST concentration shows noise and non-zero values especially in the lower edge of the tube.
The conventional method suffers from the same but weaker effect.
The values at the lower edge are expected to result from a low signal intensity due to its positioning at the open end of the head coil with vanishing coil sensitivities combined with the coronal slice orientation.
Therefore, the same effect is not expected to occur in the in vivo studies.
This problem could possibly be solved by adjusting the regularization parameters or using Total-Variation (TV) regularization, which would be ideal for a phantom with homogenous regions and sharp edges. 
However, for the sake of consistency, TGV was used for all measurements as it is more suitable for in vivo data.
For all other ROIs, which contain CEST agents, the noise does not seem to be a problem and the proposed methods replicates the conventional method.    

Validation of the proposed method showed that the saturation time influences the prediction of magnetization in the steady state, with the uncertainty decreasing as the saturation time increases.  This presents the trade-off between total acquisition time and the accuracy of the steady-state estimation depending on CEST concentration and $T_1$ values. As lower acquisition time is desired, but sufficient saturation time is needed, $T_{sat}$=1.5s was chosen as the compromise for the proposed method. The complete results for the saturation time  investigation can be found in the Supplementary Document S2. Test-retest measurements were also performed for the phantom. The results are shown in the Supplementary Document S3 and show good repeatability for the proposed method.  The differences between the CEST metrics, visible in Figure \ref{fig:res:05}, highlight the need for a careful choice of the quantitative CEST metric. 
The difference between non QUASS and QUASS corrected methods show the strong influence of the saturation.
While the QUASS correction can compensate for the saturation, the influence of $T_1$ needs to be incorporated as well.
In the QUASS $MTR_{REX}$ map the amplitudes correlate strongly with $T_1$ and without $T_1$ correction little knowledge about the underlying CEST agent concentration can be observed.
An additional correction of QUASS with $T_1$ can be achieved by further post-processing the data to calculate the $MTR_{AREX}$.
The proposed method avoids multiple post-processing steps and acquisitions providing saturation corrected CEST and $T_1$ measurements in a single scan with a reduced measurement duration.

\subsection{In vivo Measurements}
The results present a good agreement between the $T_1$ values of the conventional and proposed method. 
The increased variance for higher $T_1$ values is similar to the effect observed in-vitro, but also results from the inhomogeneous nature of the gray matter regions. 
The $T_1$ over all slices both for the IR-LL and the proposed method are consistent.
This is expected, because both use the same readout parameters and the same inversion pulse.

The CEST amplitude maps show good agreement between the two methods, with considerably more noise for the conventional method and more homogenous signal in the proposed method. 
Some central regions of the brain show higher values for the proposed method, while some parts of gray matter seem to show lower values. 
Here, the reason is not yet understood, as the same saturation is ensured and $B_1$ effects should not have any influence. Additional investigations of the relative $B_1$ and also $B_0$ inhomogeneities showed only weak correlation with the observed differences. These results are shown in the Supplementary Document S4. It will be part of future work to investigate this effect further.

In general, in vivo CEST at 3T at high resolution is challenging, due to the low SNR and the low spectral resolution.
The proposed method shows a homogenous signal over all slices, which is not the case for the conventional method.
Here, the influence of the Cartesian 3D readout could also be a source of the differences as the conventional slices show a much higher variance, which we would not expect from three slices in proximity. 
The proposed method uses a golden-ratio based stack-of-stars readout, which is more robust against motion artifacts, which could explain some differences, even though motion correction was applied to the reference data.   Furthermore, for the proposed method each measured k-space line is time encoded, leading to a more accurate reconstruction of the data. Combined with the TGV regularization, more homogenous CEST maps for all slices are achieved with higher SNR. With single-shot Cartesian encoding, magnetization changes during the measurement of a k-space, resulting in k-space filtering and blurring in the image.    This could also explain some differences between the methods.  

As seen in the phantom results, the proposed method shows very good agreement to the conventional method, while being more time efficient.
\subsection{Limits of the proposed model and other fast quantification methods}
If the model does not accurately describe the measured signal, the reconstruction will not produce accurate parameter maps. However, the same applies to quantification in the image domain. The proposed model estimates the steady-state Z-spectrum from the acquired data, which could add uncertainty. In the performed experiments, the model accurately predicted the steady-state z-spectrum, in the both the numerical phantom and the phantom measurements, with the used saturation time of 1.5s. Additionally, the signal model assumes a Lorentzian lineshape, requiring optimized OC pulses for accurate fitting, as Gaussian pulses produce non-Lorentzian shapes not well captured by the model  \cite{Huemer2023PV}\cite{Stilianu2024}. The Look-Locker readout is well established and widely validated \cite{Wang2021}. Currently, the reconstruction is performed slice-wise, neglecting correlations between slices and limited to 80 time points per offset due to memory constraints. Additional time points and 3D reconstruction could further improve the results.   

For faster CEST quantification, multiple other methods have been proposed. One approach is to reconstruct the images from undersampled k-space data first, and then perform fitting in image space \cite{Zu2024}\cite{Weigand-Whittier2023}. The undersampled acquisition saves time, but the image reconstruction and fitting are separate steps, which can lead to error propagation and loss of information compared to the proposed model-based reconstruction. Undersampling can also be done in the offset dimension \cite{Yang2024}, which reduces the acquisition time significantly, but reduces the spectral resolution. Another very promising approach is MRF-CEST \cite{Cohen2018}\cite{Cohen2023}\cite{Perlman2022}, which uses a variable saturation scheme to create unique signal evolutions for different tissue parameters. This allows for very fast acquisitions (some reports show acquisition times as low as two minutes), but the quantification is limited to one CEST pool and the MT pool. Furthermore, the dictionary resolution and the accuracy of the simulated signal evolutions limits the accuracy of the quantification.  

\subsection{Model-based reconstruction for CEST imaging}
To our knowledge this is the first method to directly quantify $R_{ex}$ and $T_1$ from a single acquisition and the first application of a model-based reconstruction for CEST imaging.
The establishment of model-based reconstruction for CEST imaging should enable many new applications, such as the use of more complex models, e.g. imaging of the saturation process by interleaved saturation and acquisition \cite{ESMRMB2024tCEST} or the incorporation of $B_0$ and $B_1$ maps into the reconstruction. 

This study clearly shows the advantages of model-based reconstruction, where all the measured k-lines contribute to the result through the underlying model.

\section{Conclusion}
In this paper, we presented a novel sequence and model-based reconstruction method for direct quantification of $R_{ex}$ and $T_1$ from a single acquisition. 
The combination of the CEST saturation and the 3D SOS FLASH readout provides a time-efficient way to acquire the Z-spectrum and $T_1$ information of multiple slices in a single measurement. 
The measurement time is reduced by 20-30$\%$ compared to the conventional CEST measurement combined with a separate $T_1$ mapping sequence.
The combined signal model can predict the steady-state Z-spectrum using the $T_1$ information gained from the Look-Locker readout, which allows for direct quantification of $R_{ex}$.
The proposed method was validated using a numerical phantom and compared to conventional CEST measurements and Look-Locker $T_1$-mapping in phantom and in vivo experiments on a 3T clinical scanner.
The sequence as well as the reconstruction methods are available as open-source software. 

\section*{Open Research}
\subsection*{Data Availability Statement}
In the spirit of reproducible research, code and data to reproduce the experiments is available at: https://doi.org/10.5281/zenodo.15768062.

PyQMRI is available on \url{https://gitlab.tugraz.at/ibi/mrirecon/software/PyQMRI}.

BART is available on \url{https://mrirecon.github.io/bart/}.

PulseqCEST is available on \url{https://pulseq-cest.github.io/}. 

\section*{Acknowledgment}
This research was funded in whole, or in part, by the Austrian Science Fund (FWF) I4870 and 10.55776/F100800. For the purpose of open access, the author has applied a CC BY public copyright license to any Author Accepted Manuscript version arising from this submission.

\clearpage
\clearpage

\clearpage
\listoffigures
\clearpage

\clearpage


\begin{thebibliography}{99}
  {}
  \bibitem{vanZijl2018}
  P. C. van Zijl, W. W. Lam, J. Xu, L. Knutsson, and G. J. Stanisz, “Magnetization transfer
  contrast and chemical exchange saturation transfer mri. features and analysis of the
  field-dependent saturation spectrum,” \emph{NeuroImage}, vol. 168, pp. 222–241, Mar.
  2018, \textsc{issn}: 10959572. \textsc{doi}:  {10.1016/j.neuroimage.2017.04.045}.
  {}
  \bibitem{Jones2018}
  K. M. Jones, A. C. Pollard, M. D. Pagel, M. Pagel, K. Jones, and A. Pollard, “Clinical
  applications of chemical exchange saturation transfer (cest) mri,” \emph{Journal of
  Magnetic Resonance Imaging}, vol. 47, pp. 11–27, 1 Jan. 2018, \textsc{issn}: 1522-2586.
  \textsc{doi}:  {10.1002/JMRI.25838}.
  {}
  \bibitem{Goerke2019}
  S. Goerke, Y. Soehngen, A. Deshmane, \emph{et al.}, “Relaxation-compensated apt and
  rnoe cest-mri of human brain tumors at 3 t,” \emph{Magnetic Resonance in Medicine},
  vol. 82, pp. 622–632, 2 Aug. 2019, \textsc{issn}: 0740-3194. \textsc{doi}: 
  {10.1002/mrm.27751}.
  {}
  \bibitem{Zhang2018}
  J. Zhang, W. Zhu, R. Tain, X. J. Zhou, and K. Cai, “Improved differentiation of low-grade
  and high-grade gliomas and detection of tumor proliferation using apt contrast fitted from
  z-spectrum,” \emph{Molecular Imaging and Biology}, vol. 20, pp. 623–631, 4 Aug. 2018,
  \textsc{issn}: 18602002. \textsc{doi}: 
  {10.1007/S11307-017-1154-Y/FIGURES/6}.
  {}
  \bibitem{Paech2018}
  D. Paech, J. Windschuh, J. Oberhollenzer, \emph{et al.}, “Assessing the predictability of
  idh mutation and mgmt methylation status in glioma patients using relaxation-compensated
  multipool cest mri at 7.0 t,” \emph{Neuro-Oncology}, vol. 20, pp. 1661–1671, 12 Nov. 2018,
  \textsc{issn}: 1522-8517. \textsc{doi}:  {10.1093/NEUONC/NOY073}.
  {}
  \bibitem{Leung2024}
  V. W. M. Leung, S. W. Park, J. H. C. Lai, \emph{et al.}, “Imaging treatment efficacy of
  repeated photodynamic therapy in glioblastoma using chemical exchange transfer saturation
  mri,” \emph{Magnetic Resonance in Medicine}, Nov. 2024, \textsc{issn}: 1522-2594.
  \textsc{doi}:  {10.1002/MRM.30362}.
  {}
  \bibitem{Abrar2021}
  D. B. Abrar, C. Schleich, K. L. Radke, \emph{et al.}, “Detection of early cartilage
  degeneration in the tibiotalar joint using 3 t gagcest imaging: A feasibility study,”
  \emph{Magnetic Resonance Materials in Physics, Biology and Medicine}, vol. 34,
  pp. 249–260, 2 Apr. 2021, \textsc{issn}: 13528661. \textsc{doi}: 
  {10.1007/S10334-020-00868-Y/FIGURES/6}.
  {}
  \bibitem{Zhou2024}
  L. Zhou, W. Pan, R. Huang, \emph{et al.}, “Amide proton transfer-weighted mri,
  associations with clinical severity and prognosis in ischemic strokes,” \emph{Journal of
  Magnetic Resonance Imaging}, vol. 60, pp. 2509–2517, 6 Dec. 2024, \textsc{issn}:
  1522-2586. \textsc{doi}:  {10.1002/JMRI.29333}.
  {}
  \bibitem{Zhou2022}
  J. Zhou, M. Zaiss, L. Knutsson, \emph{et al.}, “Review and consensus recommendations on
  clinical apt-weighted imaging approaches at 3t: Application to brain tumors,”
  \emph{Magnetic Resonance in Medicine}, vol. 88, pp. 546–574, 2 Aug. 2022, \textsc{issn}:
  1522-2594. \textsc{doi}:  {10.1002/MRM.29241}.
  {}
  \bibitem{Sun2024}
  P. Z. Sun, “Quasi-steady-state (quass) reconstruction enhances t1 normalization in apparent
  exchange-dependent relaxation (arex) analysis: A reevaluation of t1 correction in
  quantitative cest mri of rodent brain tumor models,” \emph{Magnetic Resonance in
  Medicine}, vol. 92, pp. 236–245, 1 Jul. 2024, \textsc{issn}: 1522-2594. \textsc{doi}:
   {10.1002/MRM.30056}.
  {}
  \bibitem{Kim2015}
  J. Kim, Y. Wu, Y. Guo, H. Zheng, and P. Z. Sun, “A review of optimization and
  quantification techniques for chemical exchange saturation transfer mri toward sensitive in
  vivo imaging,” \emph{Contrast Media \& Molecular Imaging}, vol. 10, pp. 163–178, 3 May
  2015, \textsc{issn}: 1555-4317. \textsc{doi}:  {10.1002/CMMI.1628}.
{}
\bibitem{Zaiss2011}
M. Zaiß, B. Schmitt, and P. Bachert, “Quantitative separation of cest effect from
magnetization transfer and spillover effects by lorentzian-line-fit analysis of z-spectra,”
\emph{Journal of Magnetic Resonance}, vol. 211, pp. 149–155, 2 Aug. 2011, \textsc{issn}:
10907807. \textsc{doi}:  {10.1016/j.jmr.2011.05.001}.
{}
\bibitem{Wu2015}
R. Wu, G. Xiao, I. Y. Zhou, C. Ran, and P. Z. Sun, “Quantitative chemical exchange
saturation transfer (qcest) mri – omega plot analysis of rf-spillover-corrected inverse cest
ratio asymmetry for simultaneous determination of labile proton ratio and exchange rate,”
\emph{NMR in Biomedicine}, vol. 28, pp. 376–383, 3 Mar. 2015, \textsc{issn}: 1099-1492.
\textsc{doi}:  {10.1002/NBM.3257}.
{}
\bibitem{Cohen2018}
O. Cohen, S. Huang, M. T. McMahon, M. S. Rosen, and C. T. Farrar, “Rapid and
quantitative chemical exchange saturation transfer (cest) imaging with magnetic resonance
fingerprinting (mrf),” \emph{Magnetic Resonance in Medicine}, vol. 80, pp. 2449–2463, 6
Dec. 2018, \textsc{issn}: 1522-2594. \textsc{doi}:  {10.1002/MRM.27221}.
{}
\bibitem{Cohen2023}
O. Cohen, V. Y. Yu, K. R. Tringale, \emph{et al.}, “Cest mr fingerprinting (cest-mrf) for
brain tumor quantification using epi readout and deep learning reconstruction,”
\emph{Magnetic Resonance in Medicine}, vol. 89, pp. 233–249, 1 Jan. 2023, \textsc{issn}:
15222594. \textsc{doi}: 
{10.1002/MRM.29448;WGROUP:STRING:PUBLICATION}.
{}
\bibitem{Zaiss2014}
M. Zaiss, J. Xu, S. Goerke, \emph{et al.}, “Inverse z-spectrum analysis for spillover-, mt-,
and t1-corrected steady-state pulsed cest-mri – application to ph-weighted mri of acute
stroke,” \emph{NMR in Biomedicine}, vol. 27, pp. 240–252, 3 Mar. 2014, \textsc{issn}:
1099-1492. \textsc{doi}:  {10.1002/NBM.3054}.
{}
\bibitem{Sun2021}
P. Z. Sun, “Quasi-steady state chemical exchange saturation transfer (quass cest)
analysis—correction of the finite relaxation delay and saturation time for robust cest
measurement,” \emph{Magnetic Resonance in Medicine}, vol. 85, pp. 3281–3289, 6 Jun.
2021, \textsc{issn}: 1522-2594. \textsc{doi}:  {10.1002/MRM.28653}.
{}
\bibitem{Vinogradov2011}
E. Vinogradov, T. C. Soesbe, J. A. Balschi, A. D. Sherry, and R. E. Lenkinski, “Pcest:
Positive contrast using chemical exchange saturation transfer,” \emph{Journal of Magnetic
Resonance}, vol. 215, pp. 64–73, Feb. 2011, \textsc{issn}: 10907807. \textsc{doi}:
 {10.1016/j.jmr.2011.12.011}.
{}
\bibitem{Jin2012}
T. Jin and S. G. Kim, “Quantitative chemical exchange sensitive mri using irradiation with
toggling inversion preparation,” \emph{Magnetic Resonance in Medicine}, vol. 68,
pp. 1056–1064, 4 Oct. 2012. \textsc{doi}:  {10.1002/MRM.24449}.
{}
\bibitem{Chen2024}
W. Chen, Z. Chen, L. Ma, Y. Wang, and X. Song, “Rapid and quantitative cest-mri
sequence using water presaturation,” \emph{Magnetic Resonance in Medicine}, 2024,
\textsc{issn}: 1522-2594. \textsc{doi}:  {10.1002/MRM.30309}.
{}
\bibitem{Zaiss2022}
M. Zaiss, T. Jin, S. G. Kim, and D. F. Gochberg, “Theory of chemical exchange saturation
transfer mri in the context of different magnetic fields,” \emph{NMR in Biomedicine},
vol. 35, e4789, 11 Nov. 2022, \textsc{issn}: 1099-1492. \textsc{doi}: 
{10.1002/NBM.4789}.
{}
\bibitem{Deichmann2005}
R. Deichmann, “Fast high-resolution t1 mapping of the human brain,” \emph{Magnetic
Resonance in Medicine}, vol. 54, pp. 20–27, 1 Jul. 2005, \textsc{issn}: 1522-2594.
\textsc{doi}:  {10.1002/MRM.20552}.
{}
\bibitem{Roeloffs2016}
V. Roeloffs, X. Wang, T. J. Sumpf, M. Untenberger, D. Voit, and J. Frahm, “Model-based
reconstruction for t1 mapping using single-shot inversion-recovery radial flash,”
\emph{International Journal of Imaging Systems and Technology}, vol. 26, pp. 254–263, 4
Dec. 2016, \textsc{issn}: 10981098. \textsc{doi}: 
{10.1002/IMA.22196;PAGE:STRING:ARTICLE/CHAPTER}.
{}
\bibitem{Wang2021}
X. Wang, Z. Tan, N. Scholand, V. Roeloffs, and M. Uecker, “Physics-based reconstruction
methods for magnetic resonance imaging,” \emph{Philosophical Transactions of the Royal
Society A}, vol. 379, 2200 Jun. 2021, \textsc{issn}: 1364503X. \textsc{doi}: 
{10.1098/RSTA.2020.0196}.
{}
\bibitem{Block2009}
K. T. Block, M. Uecker, and J. Frahm, “Model-based iterative reconstruction for radial fast
spin-echo mri,” \emph{IEEE Transactions on Medical Imaging}, vol. 28, pp. 1759–1769, 11
Nov. 2009, \textsc{issn}: 02780062. \textsc{doi}:  {10.1109/TMI.2009.2023119}.
{}
\bibitem{Sumpf2011}
T. J. Sumpf, M. Uecker, S. Boretius, and J. Frahm, “Model-based nonlinear inverse
reconstruction for t2 mapping using highly undersampled spin-echo mri,” \emph{Journal of
Magnetic Resonance Imaging}, vol. 34, pp. 420–428, 2 Aug. 2011, \textsc{issn}: 1522-2586.
\textsc{doi}:  {10.1002/JMRI.22634}.
{}
\bibitem{Maier2019}
O. Maier, J. Schoormans, M. Schloegl, \emph{et al.}, “Rapid t1 quantification from high
resolution 3d data with model-based reconstruction,” \emph{Magnetic Resonance in
Medicine}, vol. 81, pp. 2072–2089, 3 Mar. 2019, \textsc{issn}: 1522-2594. \textsc{doi}:
 {10.1002/MRM.27502}.
{}
\bibitem{Stilianu2024}
C. Stilianu, C. Graf, M. Huemer, \emph{et al.}, “Enhanced and robust contrast in cest mri:
Saturation pulse shape design via optimal control,” \emph{Magnetic Resonance in
Medicine}, vol. 92, pp. 1867–1880, 5 Nov. 2024, \textsc{issn}: 1522-2594. \textsc{doi}:
 {10.1002/MRM.30164}.
{}
\bibitem{Stilianu2025}
C. Stilianu, M. Huemer, M. Zaiss, and R. Stollberger, “Generalization of optimal control
saturation pulse design for robust and high cest contrast,” \emph{bioRxiv},
p. 2025.08.21.671490, Aug. 2025, \textsc{issn}: 2692-8205. \textsc{doi}: 
{10.1101/2025.08.21.671490}.
{}
\bibitem{Bernstein2004}
M. A. Bernstein, K. F. King, and X. J. Zhou, “Handbook of mri pulse sequences,”
\emph{Handbook of MRI Pulse Sequences}, pp. 1–1017, Sep. 2004. \textsc{doi}: 
{10.1016/B978-0-12-092861-3.X5000-6}.
{}
\bibitem{Winkelmann2007}
S. Winkelmann, T. Schaeffter, T. Koehler, H. Eggers, and O. Doessel, “An optimal radial
profile order based on the golden ratio for time-resolved mri,” \emph{IEEE Transactions on
Medical Imaging}, vol. 26, pp. 68–76, 1 Jan. 2007, \textsc{issn}: 02780062. \textsc{doi}:
 {10.1109/TMI.2006.885337}.
{}
\bibitem{Look1970}
D. C. Look and D. R. Locker, “Time saving in measurement of nmr and epr relaxation
times,” \emph{Review of Scientific Instruments}, vol. 41, pp. 250–251, 2 Feb. 1970,
\textsc{issn}: 0034-6748. \textsc{doi}:  {10.1063/1.1684482}.
{}
\bibitem{Fessler2003}
J. A. Fessler and B. P. Sutton, “Nonuniform fast fourier transforms using min-max
interpolation,” \emph{IEEE Transactions on Signal Processing}, vol. 51, pp. 560–574, 2
Feb. 2003, \textsc{issn}: 1053587X. \textsc{doi}:  {10.1109/TSP.2002.807005}.
{}
\bibitem{Bredies2010}
K. Bredies, K. Kunisch, and T. Pock, “Total generalized variation,”
\emph{https://doi.org/10.1137/090769521}, vol. 3, pp. 492–526, 3 Sep. 2010, \textsc{issn}:
19364954. \textsc{doi}:  {10.1137/090769521}.
{}
\bibitem{Knoll2011}
F. Knoll, K. Bredies, T. Pock, and R. Stollberger, “Second order total generalized variation
(tgv) for mri,” \emph{Magnetic Resonance in Medicine}, vol. 65, pp. 480–491, 2 Feb. 2011,
\textsc{issn}: 1522-2594. \textsc{doi}:  {10.1002/MRM.22595}.
{}
\bibitem{Huemer2024}
M. Huemer, C. Stilianu, O. Maier, \emph{et al.}, “Improved quantification in cest-mri by
joint spatial total generalized variation,” \emph{Magnetic Resonance in Medicine}, vol. 92,
pp. 1683–1697, 4 Oct. 2024, \textsc{issn}: 1522-2594. \textsc{doi}: 
{10.1002/MRM.30129}.
{}
\bibitem{Graf2021}
C. Graf, A. Rund, C. S. Aigner, and R. Stollberger, “Accuracy and performance analysis for
bloch and bloch-mcconnell simulation methods,” \emph{Journal of Magnetic Resonance},
vol. 329, p. 107 011, Aug. 2021, \textsc{issn}: 1090-7807. \textsc{doi}: 
{10.1016/J.JMR.2021.107011}.
{}
\bibitem{Schuenke2023}
P. Schuenke, K. Herz, Z. Zu, \emph{et al.}, “Validate your cest simulation!” In \emph{Proc
ISMRM}, 2023.
{}
\bibitem{BART}
\emph{Bart toolbox for computational magnetic resonance imaging}. \textsc{doi}:
 {10.5281/zenodo.592960}.
{}
\bibitem{Layton2017}
K. J. Layton, S. Kroboth, F. Jia, \emph{et al.}, “Pulseq: A rapid and
hardware-independent pulse sequence prototyping framework,” \emph{Magnetic Resonance
in Medicine}, vol. 77, pp. 1544–1552, 4 Apr. 2017, \textsc{issn}: 15222594. \textsc{doi}:
 {10.1002/MRM.26235,}
{}
\bibitem{Herz2021}
K. Herz, S. Mueller, O. Perlman, \emph{et al.}, “Pulseq-cest: Towards multi-site
multi-vendor compatibility and reproducibility of cest experiments using an open-source
sequence standard,” \emph{Magnetic Resonance in Medicine}, vol. 86, pp. 1845–1858, 4
Oct. 2021, \textsc{issn}: 0740-3194. \textsc{doi}:  {10.1002/mrm.28825}.
{}
\bibitem{Weinmller2025}
S. Weinmüller, J. Endres, N. Dang, R. Stollberger, and M. Zaiss, “Mr-zero meets
flash – controlling the transient signal decay in gradient- and rf-spoiled gradient echo
sequences,” \emph{Magnetic Resonance in Medicine}, vol. 93, pp. 942–960, 3 Mar. 2025,
\textsc{issn}: 15222594. \textsc{doi}:  {10.1002/MRM.30318}.
{}
\bibitem{Schuenke2017}
P. Schuenke, J. Windschuh, V. Roeloffs, M. E. Ladd, P. Bachert, and M. Zaiss,
“Simultaneous mapping of water shift and b1(wasabi)—application to field-inhomogeneity
correction of cest mri data,” \emph{Magnetic Resonance in Medicine}, vol. 77, pp. 571–580,
2 Feb. 2017, \textsc{issn}: 15222594. \textsc{doi}: 
{10.1002/MRM.26133}.
{}
\bibitem{Maier2020}
O. Maier, S. M. Spann, M. Bödenler, and R. Stollberger, “Pyqmri: An accelerated python
based quantitative mri toolbox,” \emph{Journal of Open Source Software}, vol. 5, p. 2727,
56 Dec. 2020, \textsc{issn}: 2475-9066. \textsc{doi}:  {10.21105/JOSS.02727}.
{}
\bibitem{Mennecke2022}
A. Mennecke, K. M. Khakzar, A. German, \emph{et al.}, “7 tricks for 7 t cest: Improving
the reproducibility of multipool evaluation provides insights into the effects of age and the
early stages of parkinson’s disease,” \emph{NMR in Biomedicine}, e4717, Mar. 2022,
\textsc{issn}: 0952-3480. \textsc{doi}:  {10.1002/nbm.4717}.
{}
\bibitem{Liu2023}
R. Liu, X. Wang, Z. Zhao, \emph{et al.}, “A comparative study of quantitative metrics in
chemical exchange saturation transfer imaging for grading gliomas in adults,”
\emph{Magnetic Resonance Imaging}, vol. 96, pp. 50–59, Feb. 2023, \textsc{issn}:
0730-725X. \textsc{doi}:  {10.1016/J.MRI.2022.11.008}.
{}
\bibitem{Rosenzweig2019}
S. Rosenzweig, H. C. M. Holme, and M. Uecker, “Simple auto-calibrated gradient delay
estimation from few spokes using radial intersections (ring),” \emph{Magnetic Resonance in
Medicine}, vol. 81, pp. 1898–1906, 3 Mar. 2019, \textsc{issn}: 1522-2594. \textsc{doi}:
 {10.1002/MRM.27506}.
{}
\bibitem{Buehrer2007}
M. Buehrer, K. P. Pruessmann, P. Boesiger, and S. Kozerke, “Array compression for mri
with large coil arrays,” \emph{Magnetic Resonance in Medicine}, vol. 57, pp. 1131–1139, 6
2007, \textsc{issn}: 15222594. \textsc{doi}:  {10.1002/MRM.21237,}
{}
\bibitem{Huang2008}
F. Huang, S. Vijayakumar, Y. Li, S. Hertel, and G. R. Duensing, “A software channel
compression technique for faster reconstruction with many channels,” \emph{Magnetic
Resonance Imaging}, vol. 26, pp. 133–141, 1 Jan. 2008, \textsc{issn}: 0730725X.
\textsc{doi}:  {10.1016/j.mri.2007.04.010}.
{}
\bibitem{Huemer2023PV}
M. Huemer, C. Stilianu, and R. Stollberger, “Improving the accuracy of
multipool-lorentzian fitting in cest mri by use of a pseudo-voigt lineshape for direct water
saturation,” in \emph{Proceedings of the 2023 ISMRM \& ISMRT Annual Meeting \&
Exhibition}, 3116, 2023.
{}
\bibitem{Zu2024}
T. Zu, X. Yong, Z. Dai, \emph{et al.}, “Prospective acceleration of whole-brain cest
imaging by golden-angle view ordering in cartesian coordinates and joint k-space and
image-space parallel imaging (kipi),” \emph{Magnetic Resonance in Medicine}, Nov. 2024,
\textsc{issn}: 1522-2594. \textsc{doi}:  {10.1002/MRM.30375}.
{}
\bibitem{Weigand-Whittier2023}
J. Weigand-Whittier, M. Sedykh, K. Herz, \emph{et al.}, “Accelerated and quantitative
three-dimensional molecular mri using a generative adversarial network,” \emph{Magnetic
Resonance in Medicine}, vol. 89, pp. 1901–1914, 5 May 2023, \textsc{issn}: 15222594.
\textsc{doi}:  {10.1002/MRM.29574}.
{}
\bibitem{Yang2024}
Z. Yang, D. Shen, K. W. Chan, and J. Huang, “Attention-based multioffset deep learning
reconstruction of chemical exchange saturation transfer (amo-cest) mri,” \emph{IEEE
Journal of Biomedical and Health Informatics}, vol. 28, pp. 4636–4647, 8 2024,
\textsc{issn}: 21682208. \textsc{doi}:  {10.1109/JBHI.2024.3404225}.
{}
\bibitem{Perlman2022}
O. Perlman, B. Zhu, M. Zaiss, M. S. Rosen, and C. T. Farrar, “An end-to-end ai-based
framework for automated discovery of rapid cest/mt mri acquisition protocols and molecular
parameter quantification (autocest),” \emph{Magnetic Resonance in Medicine}, vol. 87,
pp. 2792–2810, 6 Jun. 2022, \textsc{issn}: 15222594. \textsc{doi}: 
{10.1002/MRM.29173}.
{}
\bibitem{ESMRMB2024tCEST}
M. Huemer, C. Stilianu, and R. Stollberger, “Transient model based cest imaging—tcest,” in
\emph{Magnetic Resonance Materials in Physics, Biology and Medicine 2024 37:1}, vol. 37,
Springer, Sep. 2024, pp. 31–33. \textsc{doi}:  {10.1007/S10334-024-01191-6}.
\end{thebibliography}
\end{document}